\documentclass[reqno,12pt]{amsart}
\usepackage{fullpage}
\usepackage{amsfonts}
\usepackage{amssymb}
\usepackage{subcaption}
\usepackage{graphicx}
\usepackage{xcolor}

\numberwithin{equation}{section}
\def\sgn{\mathrm{sgn}}

\def\u{{\mathbf u}}
\def\T{{\mathcal T}}
\def\Lagr{{\mathcal L}}

\def\triv{\mathrm{triv}}
\def\shft{\mathrm{shift}}

\def\X{{\mathbf X}}
\def\Y{{\mathbf Y}}
\def\pr{\mathrm{pr}}
\def\P{P}
\def\solnsp{{\mathcal E}}

\def\A{\hat{A}}

\def\G{{\mathcal G}}

\def\alg#1{\mathfrak{#1}}

\def\Rnum{\mathbb{R}}

\def\sgn{{\rm sgn}}
\def\spans{{\rm span}}
\def\t{{\rm t}}

\def\smallint{\textstyle\int}

\def\Ref#1{Ref.~\cite{#1}}

\newtheorem{prop}{Proposition}
\newtheorem{thm}{Theorem}

\title{Hidden symmetry group for particle orbits\\ (timelike geodesics) in Schwarzschild spacetime}

\author{
Stephen C. Anco
\lowercase{\scshape{and}}
Mahdieh Gol Bashmani Moghadam
\\
\\\lowercase{\scshape{
Department of Mathematics and Statistics, 
Brock University\\
St. Catharines, ON, Canada}} \\
}

\begin{document}

\begin{abstract}
For the timelike geodesic equations in Schwarzschild spacetime, 
three hidden conserved quantities were found recently,
which are analogues of dynamical quantities 
related to the well-known Laplace--Runge--Lenz (LRL) vector in Newtonian gravity.
In particular, the geodesic equations possess
an LRL angle, an LRL Killing-vector time and an LRL proper time,
each of which is a (piecewise) conserved quantity for all timelike geodesics. 
The present work provides a natural symmetry interpretation 
for these three quantities by applying Noether's theorem in reverse
to the geodesic Lagrangian.
This yields three new (hidden) LRL symmetry transformations.
Unlike the Killing symmetries, 
which constitute point transformations on the geodesic configuration variables
and close without the use of the geodesic equations, 
the new symmetries are dynamical transformations
which essentially involve the geodesic momenta variables
and close only on solutions of the geodesic equations.
The commutator structure of the LRL symmetries is shown to depend on
whether a geodesic is bounded or not, and horizon-crossing or not,
while their physical meaning is connected to their action
on the geodesic energy and angular momentum.
Together with the Killing symmetries and re-parameterization symmetries of the geodesics Lagrangian, 
they generate the complete Noether symmetry group 
of the timelike geodesic equations. 
\end{abstract}

\maketitle

\section{Introduction}\label{sec:intro}

In General Relativity, 
the Schwarzschild spacetime is the unique spherically symmetric solution of
the Einstein equations in vacuum \cite{Bir}. 
Its timelike geodesics describe orbits of (test) particles
around a static, spherical black hole \cite{Mis.Tho.Whe,Wal}. 
A classification of all of the different types of orbits was first derived \cite{Hag} 
soon after the Schwarzschild solution was discovered in 1916. 
More comprehensive treatments were given in subsequent decades
(see e.g. \Ref{Dar,Mis.Tho.Whe,Cha}). 

Due to spherical symmetry,
any timelike geodesic that starts in an equatorial plane will remain in that plane,
and thus there is no loss of generality in restricting attention to the equatorial geodesics.
Two scalar conserved quantities
kinematically characterize these geodesics:
energy and angular momentum, 
which arise respectively from the time-translation Killing vector
and the rotational Killing vector in the equatorial plane.
Through Noether's theorem, these conserved quantities can also be understood
in terms of the underlying symmetries of the geodesic equations,
which has attracted substantial on-going interest from many perspectives 
\cite{Kat.Lev,Pri.Cra,Bok.Kar.Kas.Zam,Hus.Mah.Qad,Tsa.Pal,Joh.Mah.Kar.Bok.Zam,Ali.Fer.Ali,Jam.Fer,Rez-Hej,Mau.Tor,Cie.Mac}. 

Remarkably, as explained in recent work \cite{Anc.Faz},
three additional conserved quantities exist for equatorial geodesics.
They are not connected with Killing vectors
or any other geometrical structures in Schwarzschild spacetime.
Instead they are analogues of dynamical quantities 
related to the Laplace-Runge-Lenz (LRL) vector in Newtonian gravity. 

Recall that, for particle motion in Newtonian gravity,
the LRL vector lies in the plane of motion
and points towards the periapsis on a particle's orbit
\cite{Gol.Poo.Saf-book,Cor-book}. 
The angular direction of this vector, called the LRL angle,
is a conserved quantity, with the same physical and mathematical status as
the Newtonian energy and angular momentum.
The time at which the particle reaches a periapsis is also
a conserved quantity \cite{Anc.Mea.Pas},
which will be referred to as the LRL time. 
For elliptical orbits, 
the LRL time is only piecewise conserved along the orbit,
since as the particle passes through the periapsis,
this quantity will jump to the time of the next periapsis.
Hence it is multi-valued, whereas for parabolic and hyperbolic orbits,
the LRL time is single-valued. 

These quantities have a natural extension to any central force
\cite{Fra,Muk,Per},
in particular to post-Newtonian gravity with spherical symmetry 
where the inverse-square force acquires a cubic correction \cite{Mis.Tho.Whe}. 
As is well known, this correction leads to precession of elliptical orbits, 
and consequently both the LRL angle and LRL time are only piecewise conserved
for such orbits \cite{Anc.Mea.Pas,Per}. 
Nevertheless, these two quantities completely characterize
the essential dynamical features of these precessing elliptical orbits. 

More broadly,
the existence of the LRL angle and LRL time as conserved quantities for particle orbits 
depends solely on the property that equations of motion have spherical symmetry.
Thus, finding analogues of these quantities for the timelike geodesic equations
in Schwarzschild spacetime is natural, even if surprising and overlooked. 
In fact, there are two analogues of LRL time:
one will be related to the Killing-vector time, 
and the other will arise from the proper time as the geodesic parameter. 
Together with the analogue of the LRL angle,
this gives a total of three new conserved quantities for timelike geodesics. 

The purpose of the present paper is to explain mathematically that
there is a natural symmetry interpretation of
the three analogue LRL quantities for the timelike geodesics in Schwarzschild spacetime.
This interpretation comes from the geodesic Lagrangian
by applying Noether's theorem in reverse to these quantities.
The familiar form of Noether's theorem states that
every symmetry transformation of an action principle gives rise to a conserved quantity.
This correspondence can be reversed so that a symmetry transformation
can be derived from every conserved quantity. 
In particular, if a complete set of conserved quantities is known,
then the corresponding set of symmetry transformations gives
the complete Noether symmetry group.
These transformations are not necessarily point transformations and may consist of
dynamical transformations that close only on solutions of the geodesic equations.
Previous work on conserved quantities of geodesics using Noether symmetries
in spherically symmetric spacetimes \cite{Kat.Lev,Pri.Cra,Jam.Fer,Rez-Hej} 
was restricted to point symmetries and thus missed finding the new dynamical transformations. 

The dynamical symmetry transformations coming from each of the LRL quantities
comprise a one-dimensional local Lie group,
and the compositions of these groups closes up to isometries
generated by the equatorial Killing vectors,
thereby yielding a non-abelian local Lie group
with a non-trivial Lie algebra structure. 
In this group,
the angular symmetry transformations corresponding to the LRL angle 
act on the equatorial geodesics
by shifting the angular momentum while keeping the energy fixed;
the temporal symmetry transformations
in the case of the LRL Killing-vector time
act by shifting the geodesic energy
and keeping the geodesic angular momentum unchanged,
and in the case of the LRL proper time,
both the angular momentum and energy of geodesics are scaled
such that their ratio is kept fixed.

The rest of the paper is organized as follows.
Section~\ref{sec:noether}
explains a modern formulation of Noether's theorem
for the geodesic equations in an arbitrary spacetime. 
Section~\ref{sec:conserved.quantities}
provides a review of the analogue LRL conserved quantities
for the timelike equatorial geodesic equations in Schwarzschild spacetime.
Section~\ref{sec:symm.generators}
gives the derivation of the equatorial LRL-type symmetry generators 
from the geodesic Lagrangian 
and discusses their basic features.
Section~\ref{sec:commutators.symmactions}
discusses the commutators of the equatorial symmetry generators 
and works out how these generators act on the conserved quantities. 
Section~\ref{sec:transformations}
derives the symmetry group transformations produced by
the equatorial symmetry generators
and discusses their physical interpretation.
Section~\ref{sec:nonequatorial}
extends the the main results on conserved quantities and variational symmetries
to non-equatorial timelike geodesics. 
Section~\ref{sec:group}
presents the symmetry group transformations for non-equatorial timelike geodesics. 
Section~\ref{sec:conclude} makes some concluding remarks.

\section{Noether symmetry group}\label{sec:noether}

There is a general correspondence between conserved quantities and symmetries of
timelike geodesics in any spacetime,
which comes from Noether's theorem applied to the geodesic action principle
\begin{equation}\label{action}
  S[x(\lambda)] = \int_{\lambda_1}^{\lambda_2} \sqrt{-\Lagr}\, d\lambda
\end{equation}
for extremization of proper time $\tau$, with fixed endpoints.
Here 
\begin{equation}\label{Lagr}
  \Lagr =   g_{\mu\nu}(x(\lambda)) \dot{x}^\mu(\lambda)  \dot{x}^\nu(\lambda)
\end{equation}
is the Lagrangian,
$g_{\mu\nu}(x)$ is the spacetime metric in coordinates $x^\mu$
(with sign convention $-+++$), 
$\lambda$ is an arbitrary parameter,
and a dot denotes derivative with respect to $\lambda$.
The Euler-Lagrange equations of motion are given by
the vanishing of the variational derivative
\begin{equation}\label{EL.geodesic.eqns}
  0= \frac{\delta\sqrt{-\Lagr}}{\delta x^\mu} =
  \sqrt{-\Lagr}^{-1} g_{\mu\nu}(x) \big( \dot{x}^\alpha\nabla_\alpha \dot{x}^\nu -(\ln\sqrt{-\Lagr})\dot{}\, \dot{x}^\nu \big) . 
\end{equation}
Their geometric meaning is that
the tangent vector $\dot{x}^\nu$ to a geodesic $x^\mu(\lambda)$ 
remains parallel to itself under transport along the geodesic,
$\dot{x}^\alpha\nabla_\alpha \dot{x}^\nu  = (\ln\sqrt{-\Lagr})\dot{}\, \dot{x}^\nu$. 
As is well known, 
they are equivalent to the geodesic equations 
\begin{equation}\label{geodesic.eqns}
  \u^\nu\nabla_\nu \u^\mu =0
\end{equation}
in terms of the 4-velocity 
\begin{equation}\label{4vel}
  \u^\mu = \frac{dx^\mu}{d\tau} = N^{-1} \dot{x}^\mu,
\end{equation}
where 
\begin{equation}\label{lapse}
  N=\frac{d\tau}{d\lambda} = \sqrt{-\Lagr}
\end{equation}
is a lapse function which relates the parameter $\lambda$ and the proper time $\tau$.
In particular, note that
\begin{equation}\label{Lagr.id}
  g^{\mu\nu}\frac{\delta\sqrt{-\Lagr}}{\delta x^\nu}
  = \sqrt{-\Lagr}\, \u^\nu\nabla_\nu \u^\mu 
\end{equation}
holds identically from equation \eqref{EL.geodesic.eqns}
combined with relation \eqref{4vel}. 

To motivate the formulation of Noether's theorem, 
an explanation of a Killing vector as a symmetry of the Lagrangian \eqref{Lagr}
will be given first. 

Suppose the metric possesses a Killing vector $\xi^\mu(x)$,
namely $\nabla^{(\nu}\xi^{\mu)}=0$. 
As a vector field in spacetime,
the Killing vector acts on the coordinates $x^\mu$
by $\xi^\alpha \partial_\alpha x^\mu = \xi^\mu$. 
These coordinates represent half of the Lagrangian variables. 
The Killing vector field can be prolonged to act
on the other half of the Lagrangian variables $\dot{x}^\mu$ 
by requiring that the prolongation commutes with differentiation with respect to $\lambda$,
which yields 
$\big(\pr\, \xi^\alpha\partial_\alpha\big) \dfrac{dx^\mu}{d\lambda}
  = \dfrac{d \xi^\mu}{d\lambda} = \dot{x}^\alpha \partial_\alpha \xi^\mu$
where $\pr$ denotes the prolongation.
In effect,
the prolonged Killing vector field acts like a total derivative operator by chain rule,
\begin{equation}\label{KV.X}
\pr\, \xi^\alpha\partial_\alpha = \xi^\alpha\partial_{x^\alpha} +  \dot{x}^\nu \partial_\nu \xi^\alpha\partial_{\dot{x}^\alpha} . 
\end{equation}

Then the prolonged Killing vector field is found to annihilate the Lagrangian:
\begin{equation}\label{KV.Lagr}
  \big(\pr\, \xi^\alpha\partial_\alpha\big) \Lagr
  = \xi^\alpha\partial_\alpha g_{\mu\nu} \dot{x}^\mu\dot{x}^\nu
  + 2g_{\mu\nu} \partial_\alpha \xi^\mu \dot{x}^\alpha \dot{x}^\nu
  = 2\nabla_{(\mu} \xi_{\nu)} \dot{x}^\mu\dot{x}^\nu =0 .
\end{equation}
This can be seen alternatively by adapting coordinates to the Killing vector  \cite{Wal}
so that $\xi^\mu \partial_\mu =\partial_y$,
which has a trivial prolongation $\pr\, \xi^\alpha\partial_\alpha = \partial_y$
and annihilates the Lagrangian because $g_{\mu\nu}$ will be independent of $y$.

A conserved quantity arises from combining expression \eqref{KV.Lagr}
with the following variational identity
\begin{equation}\label{KV.var.identity}
  \big(\pr\, \xi^\alpha\partial_\alpha\big) \sqrt{-\Lagr}  = 
  \frac{\delta \sqrt{-\Lagr}}{\delta x^\mu} \xi^\mu
+\frac{d}{d\lambda} \Big( \frac{\partial\sqrt{-\Lagr}}{\partial \dot{x^\mu}} \xi^\mu \Big)\end{equation}
where 
\begin{equation}
\frac{\partial\sqrt{-\Lagr}}{\partial\dot{x^\mu}}
= -N^{-1} g_{\mu\nu}\dot{x}^\nu = -\u_\mu .
\end{equation}
Substitution of this identity \eqref{KV.var.identity} into the lefthand side of expression \eqref{KV.Lagr} 
gives 
\begin{equation}
  \big(\pr\, \xi^\alpha\partial_\alpha\big) \Lagr
  = 2\sqrt{-\Lagr} \Big(
  -\frac{\delta\sqrt{-\Lagr}}{\delta x^\mu} \xi^\mu
  + \frac{d}{d\lambda} \Big( \u_\mu \xi^\mu \Big)
  \Big) =0 ,
\end{equation}
yielding
\begin{equation}\label{KV.noether}
  \frac{d}{d\lambda} \Big( \u_\mu \xi^\mu \Big)
  = \frac{\delta\sqrt{-\Lagr}}{\delta x^\mu} \xi^\mu
  =0 
\end{equation}
for all geodesics.
This recovers the well-known result that $g(\u,\xi)$ is conserved.

The preceding steps hold more generally for any symmetry of the Lagrangian 
in the following general sense (see e.g.\ \Ref{Kat.Lev,Pri.Cra}). 
Consider a vector field of the form 
\begin{equation}\label{X}
  \X= P^\mu \partial_{x^\mu}
\end{equation}
in the space of Lagrangian variables $(x^\mu,\dot{x}^\mu)$,
where $P^\mu$ is a function of these variables in addition to $\lambda$. 
Such vector fields can be prolonged to act on $\dot{x}^\mu$
by the requirement 
\begin{equation}\label{prolong.cond}
  \pr\X(\dot{x}^\mu)=\frac{d}{d\lambda} \X(x^\mu)
  = \frac{dP^\mu}{d\lambda} . 
\end{equation}
This yields, by chain rule, 
\begin{equation}\label{prX}
  \pr\X = P^\mu\partial_{x^\mu}
  + (\partial_{\lambda} P^\mu + \partial_{x^\alpha} P^\mu \dot{x}^\alpha + \partial_{\dot{x}^\alpha} P^\mu \ddot{x}^\alpha) \partial_{\dot{x}^\mu} ,
\end{equation}
which can be viewed as a linear differential operator.
Note that, unlike the prolonged Killing vector field \eqref{KV.X},
in general $\pr\X$ is not a vector field in the coordinate space 
$(\lambda,x^\mu,\dot{x}^\mu)$
because second derivatives $\ddot{x}^\mu$ 
appear in its (prolonged) components.

\subsection{Variational symmetries and Noether's theorem}
A \emph{variational symmetry} is a linear differential operator \eqref{prX}
such that 
\begin{equation}\label{var.symm}
  \pr\X(\sqrt{-\Lagr}) = \frac{d\Psi}{d\lambda} 
\end{equation}
holds for some function $\Psi(\lambda,x,\dot{x})$.
This general notion of a symmetry of the Lagrangian
is equivalent to preservation of the geodesic action principle
up to end point terms under the infinitesimal transformation
\begin{equation}
  S[x(\lambda)] \to
  S[x(\lambda)] +   \varepsilon \int_{\lambda_1}^{\lambda_2} \pr\X(\sqrt{-\Lagr}) \,d\lambda
  + O(\varepsilon^2)
  = S[x(\lambda)] +   \varepsilon\Psi \Big|_{\lambda_1}^{\lambda_2} + O(\varepsilon^2)
\end{equation}
where $\varepsilon$ is the parameter of the transformation. 
In particular, every variational symmetry preserves the extremals of the action principle,
and thus the geodesic equations \eqref{geodesic.eqns} are preserved.

Any linear differential operator \eqref{prX} satisfies the variational identity 
\begin{equation}\label{var.identity}
  \begin{aligned}
  \pr\X(\sqrt{-\Lagr})
  & =   \frac{\delta\sqrt{-\Lagr}}{\delta x^\mu} \X(x^\mu)
  +\frac{d}{d\lambda} \Big( \frac{\partial\sqrt{-\Lagr}}{\partial \dot{x^\mu}} \X(x^\mu) \Big)
  \\&
  = \sqrt{-\Lagr}\, g_{\mu\nu} \X(x^\mu) \u^\alpha\nabla_\alpha \u^\nu 
  -\frac{d}{d\lambda} \Big( \u_\mu \X(x^\mu) \Big)
  \end{aligned}
\end{equation}
from which the Euler-Lagrange equations of motion are derived,
with use of identity \eqref{Lagr.id}. 

Combining equations \eqref{var.identity} and \eqref{var.symm} yields
the Noether identity
\begin{equation}\label{noether.identity}
  \frac{d}{d\lambda} \Big( \u_\mu \X(x^\mu) +\Psi \Big)
  = \sqrt{-\Lagr}\, g_{\mu\nu} \X(x^\mu) \u^\alpha \nabla_\alpha \u^\nu 
\end{equation}
holding for variational symmetries. 
When the identity \eqref{noether.identity} is evaluated for geodesics,
the right hand side vanishes,
and hence the left hand side gives a conserved quantity.
This constitutes the familiar version of Noether's theorem.

\begin{prop}\label{prop:noether}
Each variational symmetry \eqref{prX}--\eqref{var.symm}
of the geodesic Lagrangian \eqref{Lagr} 
yields a locally conserved quantity
\begin{equation}\label{C.P}
  C = \u_\mu P^\mu + \Psi
\end{equation}
satisfying $\frac{dC}{d\lambda}=0$ at least piecewise in $\lambda$ 
for all solutions of the geodesic equations \eqref{geodesic.eqns}.
\end{prop}

This quantity \eqref{C.P} will be non-trivial unless
the variational symmetry is given by
$\X(x^\mu) = P^\mu =f \u^\mu$ for some function $f(\lambda,x,\dot{x})$, 
as then the Noether identity \eqref{noether.identity} implies that 
$\frac{dC}{d\lambda} = 0$ holds identically, due to $\u_\mu \u^\mu = -1$,
whereby $C$ is trivially conserved (without use of the geodesic equations).
On geodesics $x^\mu(\lambda)$,
such variational symmetries act as
\begin{equation}\label{re-param}
  \X(x^\mu(\lambda)) = f N^{-1} \dot{x}^\mu(\lambda) 
\end{equation}
which is proportional to the tangent vector $\dot{x}^\mu(\lambda)$.
This corresponds to an arbitrary re-parameterization of the geodesics, 
$x^\mu(\lambda) \to x^\mu(\tilde\lambda)$,
where
$\tilde\lambda = \lambda + \varepsilon f N^{-1} +O(\varepsilon^2)$.

Less familiar is that a converse relationship holds between
non-trivial conserved quantities and variational symmetries
(see \Ref{Pri.Cra,Olv-book,BA-book}).
From the Noether identity \eqref{noether.identity},
apply chain rule to the lefthand side, 
$\frac{d}{d\lambda} C =
\partial_\lambda C + \dot{x}^\mu\partial_{x^\mu}C + \ddot{x}^\mu\partial_{\dot{x}^\mu}C$,
and expand the righthand side through the 4-velocity relation \eqref{4vel}, 
$\u^\alpha \nabla_\alpha \u^\nu
= N^{-2} (\dot{x}^\alpha\nabla_\alpha \dot{x}^\nu + \u^\nu \u_\mu \dot{x}^\alpha\nabla_\alpha \dot{x}^\mu )$,
by using
\begin{equation}
  \u^\alpha \nabla_\alpha N = \tfrac{1}{2} N^{-1} \u^\alpha \nabla_\alpha(-\Lagr)
  = - u_\nu \u^\alpha \nabla_\alpha \dot{x}^\nu . 
\end{equation}
Hence the Noether identity becomes 
\begin{equation}
\partial_\lambda C + \dot{x}^\mu\partial_{x^\mu}C + \ddot{x}^\mu\partial_{\dot{x}^\mu}C
  = g_{\mu\nu} P^\mu N^{-1}(\dot{x}^\alpha\nabla_\alpha \dot{x}^\nu + \u^\nu \u_\mu \dot{x}^\alpha\nabla_\alpha \dot{x}^\mu ) .
\end{equation}
Then equating the terms containing $\ddot{x}^\nu$
on both sides yields
\begin{equation}
  \partial_{\dot{x}^\nu} C = N^{-1} P^\mu g_{\mu\nu} 
\end{equation}
modulo projections $\u^\nu\u_\alpha + \delta^\nu{}_\alpha$.
This establishes the following result. 

\begin{prop}\label{prop:noether.reverse}
Each non-trivial locally conserved quantity $C(\lambda,x,\dot{x})$
of the geodesic equations \eqref{geodesic.eqns}
yields a variational symmetry \eqref{prX}--\eqref{var.symm}
of the geodesic Lagrangian \eqref{Lagr},
which has components
\begin{equation}\label{P.C}
  P^\mu   = N g^{\mu\nu} \frac{\partial C}{\partial\dot{x}^\nu}
\end{equation}
modulo terms \eqref{re-param} given by symmetries
\begin{equation}\label{X.re-param}
  \X_\triv = f(\lambda,x,\dot{x}) N^{-1} \dot{x}^\mu(\lambda) \partial_{x^\mu}
\end{equation}
corresponding to re-parameterizations of the geodesics. 
\end{prop}

Thus, 
there is a one-to-one correspondence between
non-trivial conserved quantities and variational symmetries
modulo re-parameterizations symmetries. 
To examine this correspondence more closely,
return to the conserved quantity $C=\u_\mu \xi^\mu$
arising from a Killing vector $\xi^\mu$.
The formula \eqref{P.C} directly yields 
$P^\mu = \xi^\mu$ modulo $f\u^\mu$, 
which gives the variational symmetry
\begin{equation}\label{X.KV}
  \X = \xi^\mu \partial_{x^\mu} 
\end{equation}
modulo re-parameterization symmetries. 
This linear differential operator generates
a transformation that is equivalent to a Killing isometry 
\begin{equation}\label{KV.symm}
  x^\mu(\lambda) \to x^\mu(\lambda) + \xi(x(\lambda))^\mu \varepsilon + O(\varepsilon^2) .
\end{equation}  

More generally,
any variational symmetry operator \eqref{prX}
produces a group of transformations in the space of variables
$(x^\mu,\dot{x}^\mu)$, 
analogously to how a Killing vector gives an isometry.

\begin{prop}\label{prop:symm.group}
Each variational symmetry \eqref{prX}
generates a local Lie group of symmetry transformations
defined by the exponential operator
\begin{equation}\label{X.group}
  \exp\big(\varepsilon\pr\X\big) = 1 + \sum_{n=1}^{\infty} \tfrac{1}{n!} \varepsilon^n\pr\X^n
\end{equation}
which acts on the geodesic variables $(x^\mu(\lambda),\dot{x}^\mu(\lambda))$,
where the geodesic equations are used to replace 
$\frac{d\dot{x}^\mu}{d\lambda}=\ddot{x}^\mu$
in each successive term of the Taylor series. 
Here $\varepsilon \in \Rnum$ is the group parameter, 
with $\varepsilon=0$ representing the identity. 
\end{prop}

The transformations comprising a local Lie group 
are well defined in some open neighbourhood containing the identity
(see e.g.\ \Ref{Olv-book,Ser}), namely
when the parameter $|\varepsilon|$ is sufficiently small. 
If the transformations exist globally, then a local Lie group is equivalent to
a (global) Lie group.

\subsection{Point symmetries and dynamical symmetries} 

If a variational symmetry \eqref{X} has $P^\mu(\lambda,x,\dot{x})$
being linear in $\dot{x}^\nu$ such that
\begin{equation}\label{P.point}
  P^\mu = \eta^\mu(\lambda,x) + \zeta(\lambda,x)\dot{x}^\mu
\end{equation}
in terms of some functions $\eta^\mu(\lambda,x)$ and $\zeta(\lambda,x)$, 
then the symmetry operator $\pr\X$
is called an \emph{infinitesimal point symmetry}.
The group \eqref{X.group} of transformations acting on geodesics $x^\mu(\lambda)$
will have the form 
\begin{equation}\label{point.transformation}
  x^\mu(\lambda) \to
  x^\mu(\lambda)
  + \big( \eta^\mu(\lambda,x(\lambda))
  + \zeta(\lambda,x(\lambda)) \dot{x}(\lambda) \big)\varepsilon 
  + O(\varepsilon^2) 
\end{equation}
which can be shown to be equivalent to a point transformation acting
on the variables $(\lambda,x^\mu)$:
\begin{equation}
  \lambda \to \lambda - \zeta(\lambda,x)\varepsilon + O(\varepsilon^2),
  \quad
  x^\mu \to x^\mu + \eta^\mu(\lambda,x)\varepsilon + O(\varepsilon^2)
\end{equation}
where the minus sign originates from viewing
$x^\mu(\lambda)$ as a surface in the space $(\lambda,x^\mu)$. 
Note that point transformations close without the use of the geodesic equations
and thus describe kinematical symmetry features of geodesics.
The main examples, discussed earlier, are 
Killing-vector symmetries \eqref{KV.symm}. 
(See e.g.\ \Ref{Olv-book,BA-book} for a comprehensive treatment of
point transformations and symmetries.)
These symmetry transformations comprise a (global) Lie group. 

When $P^\mu(\lambda,x,\dot{x})$ has a more general dependence on $\dot{x}^\nu$
than the linear form \eqref{P.point},
then the symmetry operator $\pr\X$
is called an \emph{infinitesimal dynamical symmetry}.
Specifically, these symmetries are characterized by the condition 
\begin{equation}\label{P.dyn}
  \frac{\partial P^\mu(\lambda,x,\dot{x})}{\partial \dot{x}^\alpha}
  \neq \delta^\mu{}_\alpha\, \zeta(\lambda,x) . 
\end{equation}
The resulting transformations \eqref{X.group} act on geodesics by 
\begin{equation}\label{dyn.transformation}
  x^\mu(\lambda) \to
  x^\mu(\lambda)
  + P^\mu(\lambda,x(\lambda),\dot{x}(\lambda))\varepsilon
  + O(\varepsilon^2) .
\end{equation}
They will close only with the use of the geodesic equations
and cannot be described as a point transformation on any finite space of variables. 
(See e.g.\ \Ref{BA-book} for a discussion of non-point transformations and symmetries.)
These transformations will be seen to comprise a local Lie group. 
All of the variational symmetries arising from the analogue LRL quantities
will be of this type.

\subsection{Symmetry actions, commutators, Lie algebra structure}\label{sec:structure}

A variational symmetry $\X$ acts as a linear differential operator \eqref{prX} 
on any locally conserved quantity $C(\lambda,x,\dot{x})$. 
After use of the geodesic equations \eqref{geodesic.eqns},
the result will be a conserved quantity, since
$\tfrac{d}{d\lambda}(\pr\X C)|_\solnsp = \pr\X(\tfrac{dC}{d\lambda})|_\solnsp= 0$
holds on the solution space $\solnsp$ of geodesics $(x^\mu(\lambda),\dot{x}^\mu(\lambda))$.
Thus,
\begin{equation}\label{prX.C}
  C'(\lambda,x,\dot{x}) := (\pr\X C)|_\solnsp
\end{equation}
satisfies $\dot{C'}|_\solnsp =0$ when $\dot{C}|_\solnsp=0$. 

Symmetry actions \eqref{prX.C} obey the property 
\begin{equation}\label{prX.C1.C2}
(\pr\X_{(C_1)} C_2)|_\solnsp = -(\pr\X_{(C_2)} C_1)|_\solnsp
\end{equation}
for any pair of locally conserved quantities $C_1$ and $C_2$,
where $\X_{(C)}$ denotes the variational symmetry
given by the Noether correspondence \eqref{P.C}
for a locally conserved quantity $C(\lambda,x,\dot{x})$. 
This is an immediate consequence of a general result \cite{Anc.2026,Olv-book,BA-book}
for dynamical Lagrangian systems. 
The property \eqref{prX.C1.C2} shows that
the action of a variational symmetry operator $\X_{(C)}$ vanishes on $C$ itself, 
\begin{equation}\label{prX.C.C}
  \pr\X_{(C)} C|_\solnsp =0 . 
\end{equation}
A related property is the chain rule
\begin{equation}\label{prX.chainrule}
  \pr\X_{(C_1)} f(C_2) = f'(C_2)\pr\X_{(C_1)} C_2, 
  \quad
  \pr\X_{(f(C_1))} f(C_2) = f'(C_1)\pr\X_{(C_1)} C_2 . 
\end{equation}

Another general result \cite{Olv-book,Anc.2026} for dynamical Lagrangian systems
shows that symmetry actions
encode commutators of variational symmetries: 
\begin{equation}\label{C1.C2.commutator}
  [\pr\X_{(C_1)},\pr\X_{(C_2)}]|_\solnsp = \pr\X_{(C_3)}|_\solnsp,
  \quad
  C_3 = (\pr\X_{(C_1)} C_2)|_\solnsp = -(\pr\X_{(C_2)} C_1)|_\solnsp
\end{equation}
holding on the solution space $\solnsp$. 
However, because the geodesic Lagrangian is re-parameterization invariant,
this result turns out to hold only modulo re-parameterization symmetries \eqref{X.re-param}
when it is applied to the geodesic equations viewed as a dynamical system. 

The set of all variational symmetries of the geodesic equations forms a Lie algebra,
modulo re-parameterization symmetries. 
Since any function of conserved quantities is itself a conserved quantity,
this Lie algebra will be infinite dimensional. 
The subset of variational symmetries that are point symmetries \eqref{P.point}
comprises a Lie subalgebra that is finite dimensional,
while the remaining variational symmetries will be dynamical symmetries \eqref{P.dyn}.

In general,
it is useful to consider a maximal set of functionally independent conserved quantities,
namely,
any conserved quantity is given by some function of conserved quantities in the set. 
The variational symmetries corresponding to such a set 
need not have closed commutators. 
Although each commutator corresponds to variational symmetry arising from some conserved quantity,
these conserved commutator-quantities may be functions of 
the given set of conserved quantities.
In such situations,
the variational symmetries generate a Poisson algebra
rather than a finite-dimensional Lie algebra. 
Indeed, only when all of the conserved commutator-quantities are linear combinations of
a given set of conserved quantities will the variational symmetries generate
a finite dimensional Lie algebra, 
in which case its dimension is just equal to the maximal number of
functionally independent conserved quantities.

The same considerations apply to the local transformation groups \eqref{X.group}
generated by the set of variational symmetries.
Composition of these transformation groups forms a multi-parameter local Lie group,
whose structure is determined by the Lie algebra of all variational symmetries
\cite{Olv-book} modulo re-parameterization symmetries. 
This yields the \emph{Noether symmetry group} of the geodesic equations,
which in general will be infinite dimensional.

\section{Review of (new) conserved quantities}\label{sec:conserved.quantities}

The Schwarzschild metric
\begin{equation}\label{metric}
ds^2 = -\frac{r-2M}{r}\, dt^2 + \frac{r}{r-2M}\, dr^2 + r^2(\sin^2\theta\, d\phi^2 + d\theta^2) 
\end{equation}
exhibits spherical symmetry,
where $\phi$ is the longitude angle and $\theta$ is the colatitude angle. 
Timelike geodesics $(t(\tau),r(\tau),\phi(\tau),\theta(\tau))$ are the solutions of
the spherically-symmetric equations \eqref{geodesic.eqns}: 
\begin{subequations}\label{gen.geodesic.eqns}
  \begin{align}
    \frac{d^2t}{d\tau^2} & =
    -\frac{2M}{r(r-2M)}\frac{dt}{d\tau}\frac{dr}{d\tau} , 
    \\
    \frac{d^2r}{d\tau^2} & =
    -\frac{M(r-2M)}{r^3}\Big(\frac{dt}{d\tau}\Big)^2
    + \frac{M}{r(r-2M)}  \Big(\frac{dr}{d\tau}\Big)^2
    + (r-2M) \Big( \sin^2\theta\Big(\frac{d\phi}{d\tau}\Big)^2 +\Big(\frac{d\theta}{d\tau}\Big)^2\Big),
    \\
    \frac{d^2\phi}{d\tau^2} & =
    -\frac{2}{r}\frac{dr}{d\tau}\frac{d\phi}{d\tau}
    -2\cot\theta \frac{d\theta}{d\tau}\frac{d\phi}{d\tau},
    \\
    \frac{d^2\theta}{d\tau^2} & =
    -\frac{2}{r} \frac{dr}{d\tau} \frac{d\theta}{d\tau} 
    + \sin\theta\cos\theta \Big(\frac{d\phi}{d\tau}\Big)^2 . 
  \end{align}
\end{subequations}
This system has the familiar geometrical content that the geodesic 4-velocity
$\u = \frac{dt}{d\tau}\partial_t + \frac{dr}{d\tau}\partial_r +\frac{d\phi}{d\tau}\partial_\phi +\frac{d\theta}{d\tau}\partial_\theta$
has vanishing 4-acceleration, 
$\nabla_\u \u =0$, 
where $\nabla_\u$ 
is the covariant derivative along the geodesic.
Test particles will thereby follow timelike geodesics. 

As is well known, 
every timelike geodesic lies in an invariable plane \cite{Mis.Tho.Whe,Cha} 
which, by spherical symmetry of the Schwarzschild metric,
can be taken as the equatorial plane $\theta=\frac{1}{2}\pi$.
The resulting equatorial geodesics $(t(\tau),r(\tau),\phi(\tau),\frac{1}{2}\pi)$
are solutions of the reduced equations 
\begin{subequations}\label{equator.geodesic.eqns}
  \begin{align}
    \frac{d^2t}{d\tau^2} & =
    -\frac{2M}{r(r-2M)}\frac{dt}{d\tau}\frac{dr}{d\tau} , 
    \\
    \frac{d^2r}{d\tau^2} & =
    -\frac{M(r-2M)}{r^3}\Big(\frac{dt}{d\tau}\Big)^2
    + \frac{M}{r(r-2M)}  \Big(\frac{dr}{d\tau}\Big)^2
    + (r-2M) \Big(\frac{d\phi}{d\tau}\Big)^2 , 
    \\
    \frac{d^2\phi}{d\tau} & =
    -\frac{2}{r}\frac{dr}{d\tau}\frac{d\phi}{d\tau} . 
  \end{align}
\end{subequations}

For the purpose of understanding the symmetries that underlie these equatorial geodesics, 
it will be highly useful to view the geodesic equations \eqref{equator.geodesic.eqns}
as a dynamical system for the spacetime motion of a test particle. 
This dynamics has three degrees of freedom: 
$(t(\tau),r(\tau),\phi(\tau))$.
As a consequence of general results in dynamical systems theory \cite{Whi.Wat-book,Arn-book}, 
there will exist six \emph{locally conserved quantities}
$C(\tau,t,r,\phi,\frac{dt}{d\tau},\frac{dr}{d\tau},\frac{d\phi}{d\tau})$ satisfying
\begin{equation}\label{C}
  \frac{dC}{d\tau} =0
\end{equation}
piecewise in $\tau$ for all equatorial geodesics.
If conservation holds for all $\tau$, then $C$ is said to be \emph{globally conserved}. 

Two of the conserved quantities
arise from Killing vectors 
\begin{equation}\label{equator.KVs}
  \partial_t,
  \quad
  \partial_\phi 
\end{equation}
of the equatorial metric. 
They respectively give $C = g(\partial_t,\u)= -E$ and $C=g(\partial_\phi,\u)=L$, 
where $g$ is the Schwarzschild metric \eqref{metric},
which yields energy 
\begin{equation}\label{ener}
  E = (1 - 2M/r) \frac{dt}{d\tau} 
\end{equation}
and angular momentum
\begin{equation}\label{anglmom}
  L = r^2 \frac{d\phi}{d\tau} .
\end{equation}
A third quantity is simply the norm of the 4-velocity
\begin{equation}
  \u = \frac{dt}{d\tau}\partial_t + \frac{dr}{d\tau}\partial_r +\frac{d\phi}{d\tau}\partial_\phi
\end{equation}
for equatorial geodesics, 
\begin{equation}
-1 = -\frac{r-2M}{r} \Big(\frac{dt}{d\tau}\Big)^2 + \frac{r}{r-2M} \Big(\frac{dr}{d\tau}\Big)^2 + r^2 \Big(\frac{d\phi}{d\tau}\Big)^2 .
\end{equation}
These three physically familiar quantities are globally conserved. 
The remaining three conserved quantities come from noticing that 
the equatorial geodesic equations \eqref{equator.geodesic.eqns}
constitute a relativistic central force system.

Recall from Section~\ref{sec:intro} that 
every central force system possesses a locally conserved quantity
which is a generalization of the LRL vector for non-circular orbits in Newtonian gravity. 
The Newtonian LRL vector lies in the plane of motion
and points from the focus to the periapsis of the orbit,
and hence the angular direction of this vector is an extra constant of motion.
In addition to this quantity, the time at which a particle reaches the periapsis
represents an extra integral of motion, which depends explicitly on the Newtonian time.
It is constant for hyperbolic and parabolic orbits.
For elliptical orbits, it is multi-valued, 
since when a particle reaches an up-coming periapsis,
the LRL time jumps to the time of the next periapsis. 
This temporal quantity generalizes to any central force \cite{Anc.Mea.Pas}.
In post-Newtonian gravity with spherical symmetry, 
both the generalized LRL angle and LRL time are multi-valued
whenever an orbit is precessing \cite{Anc.Mea.Pas,Per}
(namely, they are only locally conserved). 

The generalized LRL quantities for the equatorial geodesic equations \eqref{equator.geodesic.eqns}
consist of \cite{Anc.Faz} an analogue of the Newtonian LRL angle 
\begin{equation}\label{Phi}
  \Phi = \phi - L \int_{r_0}^{r} \frac{dr}{r^2 v} , 
\end{equation}
and two analogues of the LRL time
\begin{equation}\label{T}
 T = t - E \int_{r_0}^{r} \frac{r\, dr}{(r-2M) v}
\end{equation}
and
\begin{equation}\label{Tau}
\T = \tau - \int_{r_0}^{r} \frac{dr}{v},
\end{equation}
where 
\begin{equation}\label{v}
v = \frac{dr}{d\tau} =\sgn\Big(\frac{dr}{d\tau}\Big)\sqrt{E^2-(1- 2M/r)(1+L^2/r^2)}
\end{equation}
is the radial speed.
Each of these quantities is locally conserved:
\begin{equation}
  \frac{d\Phi}{d\tau}=0,
  \quad
  \frac{dT}{d\tau}=0,
  \quad
  \frac{d\T}{d\tau}=0
\end{equation}
hold piecewise in $\tau$ for all geodesics.

Importantly, here $r_0$ is not an initial value for geodesics.
Instead it is a dynamically distinguished radial position that depends only on
the intrinsic features of the geodesics. 
Specifically,
$r_0$ will be chosen as a \emph{turning point} or a \emph{centripetal point}
or a \emph{horizon-crossing point}. 

A turning point (TP) is where the radial variable $r(\tau)$
goes through a local minimum or maximum along a geodesic,
as defined by the condition $v=\frac{dr}{d\tau}=0$.
Thus, from the radial speed equation \eqref{v}, 
these points are the (real) roots $r=r_*>2M$ of the cubic polynomial 
\begin{equation}\label{tp.eqn}
(E^2 -1) r_*^3 +2M r_*^2 -L^2 r_* +2M L^2=0 .
\end{equation}
Similarly, a centripetal point (CP) is where the radial speed $v(\tau)$
goes through a local extremum,
which is defined by $\frac{dv}{d\tau}=\frac{d^2r}{d\tau} =0$.
This point is the (real) root $r=r^*>2M$ of the quadratic polynomial 
\begin{equation}\label{cp.eqn}
M r^*{}^2 -L^2 r^*+3L^2 M=0 , 
\end{equation}
which comes from the $\tau$ derivative of the radial speed equation \eqref{v}.
A horizon-crossing point (HP) is simply where a geodesic crosses the horizon,
$r=2M$. 

Every equatorial geodesic possesses at least one of these three types of dynamically distinguished points.
(See \Ref{Cha} for a modern comprehensive discussion of the various kinds of geodesics
and their main features.)
A classification of geodesics based on turning points, centripetal points, and horizon-crossing points is given in \Ref{Anc.Faz}, 
which is reproduced here in Tables~\ref{orbits-bounded} and ~\ref{orbits-unbounded}.
The following notation is used: 
\begin{equation}\label{E_pm}
\bar{L} = L/(2M) ,
  \quad
E_\pm^2 = \tfrac{2}{3}+\tfrac{2}{27}\bar{L}^2\big( 1 \pm \sqrt{(1-3/\bar{L}^2)^3} \big) ,
\end{equation}
which have the respective ranges
$0<|\bar{L}|<\infty$; 
$\tfrac{8}{9}\leq E_-^2 <1$ and $\tfrac{8}{9}\leq E_+^2 <\infty$,
with $3\geq \bar{L}^2<\infty$.
A turning point exists only for $E<1$ when $|\bar{L}| <2$,
and for $E\leq E_+\geq 1$ when $|\bar{L}| \geq 2$; 
a centripetal point exists only for $|\bar{L}| \geq \sqrt{3}$.
(Note that in \Ref{Anc.Faz} the name ``inertial point'' was used for a centripetal point.)

\begin{table}[ht!]
\centering
\caption{
Types of bounded non-radial equatorial geodesics.
(Superscripts indicate the multiplicity of choices)
}
\label{orbits-bounded} 
\begin{tabular}{l||c|c|c}
\hline
Orbit type & $\bar{L}^2$ & $E^2$ & \parbox{1in}{\centering Distinguished\\ radial points}
\\
\hline\hline
horizon-crossing
& 
$<3$
& 
$<1$
&
TP, 
HP
\\
&
$>3$
& 
$<E_-^2$
& 
TP,
HP
\\
& 
$>3$
& 
$\geq E_-^2$ and $<E_+^2$
& 
TP,
HP
\\
&
$\geq3$ and $<4$
& 
$>E_+^2$ and $<1$
& 
TP,
CP$^2$,
HP
\\
\hline
asymptotic circular
& 
$\geq3$ 
& 
$E_+^2$
& 
HP
\\
horizon-crossing
&
&
&
\\
\hline
asymptotic circular
& 
$>3$ and $<4$
& 
$E_+^2$
& 
TP, 
CP
\\
\hline
elliptic-like
&
$>3$ and $<4$
&
$>E_-^2$ and $<E_+^2$
&
TP$^2$, 
CP
\\
&
$\geq 4$
&
$>E_-^2$ and $<1$
&
TP$^2$, 
CP
\\
\hline
\end{tabular}
\end{table}

\begin{table}[ht!]
\centering
\caption{
Types of unbounded non-radial equatorial geodesics. 
(Superscripts indicate the multiplicity of choices)
}
\label{orbits-unbounded} 
\begin{tabular}{l||c|c|c}
\hline
Orbit type & $\bar{L}^2$ & $E^2$ & \parbox{1in}{\centering Distinguished\\ radial points}
\\
\hline\hline
horizon-crossing
& 
$<3$
& 
$\geq 1$
& 
HP
\\
& 
$\geq3$ and $<4$
& 
$>1$
& 
CP$^2$, 
HP
\\
& 
$\geq4$
& 
$>E_+^2$
& 
CP$^2$, 
HP
\\
\hline
asymptotic circular
& 
$4$
& 
$1$
& 
CP
\\
parabolic-like 
&&&
\\
\hline
asymptotic circular
& 
$>4$
& 
$E_+^2$
& 
CP
\\
hyperbolic-like
&&&
\\
\hline
parabolic-like 
& 
$>4$
& 
$1$
& 
TP, 
CP
\\
\hline
hyperbolic-like
& 
$>4$
& 
$>1$ and $<E_+^2$
& 
TP, 
CP
\\
\hline
\end{tabular}
\end{table}

For geodesics that never enter the horizon,
the ones that are bounded --- elliptic-like and asymptotically circular ---
possess both a turning point and a centripetal point,
while the unbounded ones --- asymptotic-circular hyperbolic-like and parabolic-like ---
possess only a centripetal point.
In contrast, for geodesics that cross the horizon,
some possess only a turning point or a centripetal point,
some have both, and the others have neither. 
The three locally conserved LRL quantities \eqref{Phi}, \eqref{T}, \eqref{Tau}
can be evaluated explicitly in terms of elementary and elliptic functions 
for each kind of geodesic (see Appendix A in \Ref{Anc.Faz}). 

To understand the physical meaning of these quantities, 
it is useful to consider the spatial orbits given by
the projection of the equatorial geodesics into the spatial hypersurfaces
$\Sigma$ on which $t$ is constant. 
Note that, in each hypersurface $\Sigma$,
$\phi$ represents the angular direction of radial lines outward from the horizon,
and $r$ represents the position on these radial lines. 

If a geodesic possesses a turning point,
it physically describes a periapsis or an apoapsis of the orbit.
The angular quantity $\Phi$, with $r_0=r_*$,
is the angle of this point on the orbit;
the temporal quantities $T$ and $\T$, with $r_0=r_*$,
are respectively the coordinate time and the proper time
at which this point is reached.
These quantities will be multi-valued when a geodesic describes
precessing elliptic-like orbits. 
In the special case when an elliptic-like orbit is non-precessing,
$\Phi$ is single valued, while $T$, $\T$ are multi-valued. 
In all non elliptic-like orbits, the three quantities are single valued. 

If a geodesic possesses a centripetal point,
it has the physical meaning that the radial Doppler effect
(as measured by observers at spatial infinity) vanishes at this point. 
The angular quantity $\Phi$, with $r_0=r_*$,
is the angle of the centripetal point on the orbit;
the temporal quantities $T$ and $\T$, with $r_0=r_*$,
are respectively the coordinate time and the proper time at which this point is reached. 
Geodesics that do not enter the horizon have at most one centripetal point,
and so these quantities are single valued,
whereas all horizon crossing geodesics have two,
and consequently each quantity is double-valued. 

In all cases,
the LRL quantities $\Phi$, $T$, $\T$ depend only on the values of $E$ and $L$
describing the geodesic.

\section{Variational symmetries of the equatorial geodesics}\label{sec:symm.generators}

For the equatorial geodesic equations \eqref{equator.geodesic.eqns},
the geodesic Lagrangian is given by 
\begin{equation}\label{equator.Lagr}
  \Lagr = 
    {-}\frac{r-2M}{r} \dot{t}^2 + \frac{r}{r-2M} \dot{r}^2 + r^2 \dot{\phi}^2
\end{equation}
where $t$, $r$, $\phi$ are functions of an arbitrary parameter $\lambda$,
and a dot denotes the derivative with respect to $\lambda$.
The Euler-Lagrange equations of motion for $(t(\lambda),r(\lambda),\phi(\lambda))$
given by the action principle \eqref{action} 
consist of equations \eqref{equator.geodesic.eqns}
multiplied by the factor $\sqrt{-\Lagr}$, as shown by the relation \eqref{Lagr.id}.
Note that the proper time is given by $\tau = \int N\, d\lambda$,
where $N=\sqrt{-\Lagr}$ is the lapse function. 

Through the Noether correspondence \eqref{P.C} stated in Proposition~\ref{prop:noether.reverse}, 
each non-trivial locally conserved quantity
$C(\tau,t,r,\phi,\frac{dt}{d\tau},\frac{dr}{d\tau},\frac{d\phi}{d\tau})$,
satisfying $\frac{dC}{d\tau}=0$ at least piecewise in $\tau$ for all equatorial geodesics,
yields a variational symmetry of the Lagrangian \eqref{equator.Lagr}.
The explicit form of the symmetry is a prolonged linear differential operator
\begin{equation}\label{equator.prX}
  \pr\X_{(C)} =
  P^t_{(C)} \partial_t + P^r_{(C)}\partial_r + P^\phi_{(C)}\partial_\phi
  +\dot{P}^t_{(C)} \partial_{\dot{t}} + \dot{P}^r_{(C)}\partial_{\dot{r}} + \dot{P}^\phi_{(C)}\partial_{\dot{\phi}}
\end{equation}
given by 
\begin{equation}\label{equator.P.C}
  P^t_{(C)} = -\frac{r}{r-2M} N^{-1}  \frac{\partial \tilde{C}}{\partial \dot{t}}, 
  \quad
  P^r_{(C)} = \frac{r-2M}{r} N^{-1} \frac{\partial \tilde{C}}{\partial \dot{r}}, 
  \quad
  P^\phi_{(C)} =  \frac{1}{r^2} N^{-1} \frac{\partial \tilde{C}}{\partial \dot{\phi}}
\end{equation}
with
\begin{equation}\label{equator.tilC}
 \tilde C  = C(\smallint N\,d\lambda,t,r,\phi,N^{-1}\dot{t},N^{-1}\dot{r},N^{-1}\dot{\phi})
\end{equation}
expressed as a function of the Lagrangian variables and the geodesic parameter. 
In particular,
the symmetry operator \eqref{equator.prX}
induces an infinitesimal change in the Lagrangian
\begin{equation}\label{equator.prX.Lagr}
  \pr\X_{(C)}\sqrt{- \Lagr} =
  \frac{d}{d\lambda}\Big( C + 
  \frac{r-2M}{r} \u^t P^t_{(C)} - \frac{r}{r-2M} \u^r P^r_{(C)} -  r^2 \u^\phi P^\phi_{(C)} \Big) 
\end{equation}
which comes from combining equations \eqref{var.symm} and \eqref{C.P},
where
\begin{subequations}\label{equator.4vel}
  \begin{align}
    \u^t & = \frac{dt}{d\tau} = - rE/(r-2M),
    \\
    \u^r & = \frac{dr}{d\tau} = v = \sgn\Big(\frac{dr}{d\tau}\Big)\sqrt{E^2-(1- 2M/r)(1+L^2/r^2)} ,
    \label{equator.v}
    \\
    \u^\phi & = \frac{d\phi}{d\tau} = L/r^2 
  \end{align}
\end{subequations}
are the non-zero components of the 4-velocity in the equatorial plane. 
This implies that the equatorial geodesic action principle is unchanged up to end point terms,
whereby the Euler-Lagrange equations themselves are preserved.

The aim now is to obtain the variational symmetries 
arising from the complete set of five non-trivial conserved quantities
$L$, $E$, $\Phi$, $T$, $\T$
for the equatorial geodesic equations.

To start,
the angular momentum $L$ and energy $E$
given by expressions \eqref{ener} and \eqref{anglmom}, 
which arise from the Killing vectors \eqref{equator.KVs}, 
each yield a variational symmetry \eqref{equator.prX}.
These have the respective components 
$P^t_{(L)} = L \u^{t}$, $P^r_{(L)} = L \u^{r}$, $P^\phi_{(L)} = 1+L\u^{\phi}$
and
$P^r_{(E)} = -1 + E\u^{t}$, $P^r_{(E)} = E\u^{r}$, $P^\phi_{(E)} =E\u^{\phi}$ 
from relations \eqref{equator.P.C},
giving the linear differential operators 
\begin{equation}\label{equator.X.L.E}
  \X_{(L)} = \partial_\phi + L \X_\shft
  \quad
  \X_{(E)} =   -\partial_t  + E \X_\shft
\end{equation}
which act on the Lagrangian variables $(t,r,\phi)$.
Here
\begin{equation}\label{equator.shift}
  \X_\shft = \u^t\partial_t +\u^r\partial_r +\u^\phi \partial_\phi
\end{equation}
represents the generator of a shift in the proper time,
which is a particular re-parameterization of the geodesics. 

The resulting symmetry transformation groups \eqref{X.group}
generated by the variational symmetries \eqref{equator.X.L.E},
modulo re-parameterizations, 
are respectively given by 
\begin{equation}
  (t^\dagger(\tau),r^\dagger(\tau),\phi^\dagger(\tau))
  =  (t(\tau),r(\tau),\phi(\tau)+\varepsilon)
\end{equation}  
and
\begin{equation}
  (t^\dagger(\tau),r^\dagger(\tau),\phi^\dagger(\tau))
  =  (t(\tau) -\varepsilon,r(\tau),\phi(\tau)) . 
\end{equation}
Both of these groups comprise
point transformations which close independently of the equatorial geodesic equations
and coincide with the isometry groups 
generated by the Killing vectors \eqref{equator.KVs} in the equatorial plane, 
with $\varepsilon$ being the affine parameter of the respective isometries.

\subsection{LRL variational symmetries}
The variational symmetries that underlie the three LRL quantities
$\Phi$, $T$, $\T$ will be derived next.
For simplicity of subsequent expressions, 
write $\bar{L} = L/(2M)$, $\bar r=r/(2M)$,
which are dimensionless, 
and change variables from $r$ to $u=1-1/\bar r$, with $0\leq u<1$.
Let  $\nu = v|_{u=y}$ and $\nu_0=v|_{u=u_0}$,
where $v$ is the radial speed \eqref{v},
and $u_0=1-1/\bar r_0$. 
Hereafter, $u_0$ will be taken to be an arbitrary function of $\bar{L},E$.
Note that derivatives with respect to
$\dot{t}$, $\dot{r}$, $\dot{\phi}$
will appear only through $L$ and $E$ via the chain rule.

For the LRL angle $\Phi$, given by expression \eqref{Phi}:
\begin{equation}\label{XtilPhi}
\X_{(\Phi)} = \P^t_{(\Phi)}\partial_t + \P^r_{(\Phi)}\partial_r + \P^\phi_{(\Phi)}\partial_\phi
\end{equation}  
has components 
\begin{subequations}\label{X.tilPhi.components}
\begin{align}
   &\begin{aligned}
\P^t_{(\Phi)} & = 
\frac{\bar{L}}{u} \Big(
\int_{u_0}^{u} \frac{ (E^2-u) \partial_E \nu + E (\bar{L} \partial_{\bar{L}} \nu - \nu)}{\nu^2} \,dy
+\frac{(E^2 -u) \partial_E u_0 + E\bar{L} \partial_{\bar{L}} u_0}{\nu_0}
\Big),
    \end{aligned}
  \label{X.tilPhi.t}
  \\
    &\begin{aligned}
\P^r_{(\Phi)} & = 
\bar{L} v \Big(
\int_{u_0}^{u} \frac{E\partial_E \nu+\bar{L}\partial_{\bar{L}} \nu - \nu}{\nu^2}\,dy
+\frac{E\partial_E u_0+\bar{L}\partial_{\bar{L}} u_0}{\nu_0}
\Big),
   \end{aligned}
  \label{X.tilPhi.r}
  \\
    &\begin{aligned}
\P^\phi_{(\Phi)} =  
\frac{1}{2M} \Big( & 
\int_{u_0}^{u} \frac{(u-1)^2\bar{L}^2 E\partial_{E} \nu + (1 + (u-1)^2\bar{L}^2)(\bar{L}\partial_{\bar{L}} \nu - \nu)}{\nu^2} \, dy 
\\&\qquad
+ \frac{(u-1)^2\bar{L}^2E\partial_{E} u_0 +(1 + (u-1)^2\bar{L}^2)\bar{L}\partial_{\bar{L}} u_0}{\nu_0} 
\Big) .
   \end{aligned}
  \label{X.tilPhi.phi}
\end{align}
\end{subequations}

For the LRL time $T$, given by expression \eqref{T}:
\begin{equation}\label{XtilT}
  \X_{(T)} = \P^t_{(T)}\partial_t + \P^r_{(T)}\partial_r + \P^\phi_{(T)}\partial_\phi
\end{equation}    
has components
\begin{subequations}\label{X.tilT.components}
  \begin{align}
  &\begin{aligned}
  \P^t_{(T)} = 
  \frac{2 M}{u} \Big(
  \int_{u_0}^{u} \frac{E^2\bar{L} \partial_{\bar{L}} \nu + (E^2 - u)(E\partial_E \nu - \nu)}{(y-1)^2 y \nu^2} \, dy
  + \frac{E^2 \bar{L}\partial_{\bar{L}} u_0  +(E^2 - u)E\partial_E u_0}{(u_0 -1)^2 u_0\nu_0}
  \Big),
  \end{aligned}
\\
  &\begin{aligned}
\P^r_{(T)} = 
2 ME v \Big( 
\int_{u_0}^{u} \frac{\bar{L}\partial_{\bar{L}}\nu + E\partial_E\nu - \nu}{(y-1)^2 y \nu^2} \, dy
+\frac{\bar{L}\partial_{\bar{L}} u_0 + E\partial_E u_0}{(u_0-1)^2 u_0 \nu_0}
\Big) ,
   \end{aligned}
\\
 &\begin{aligned}
\P^\phi_{(T)} =  
E (u^2-1) \Big( &
\int_{u_0}^{u} \frac{(1 + (u-1)^2\bar{L}^2) \partial_{\bar{L}} \nu+ (u-1)^2\bar{L}(E\partial_E \nu - \nu)}{(y-1)^2 y \nu^2} \, dy
\\&\qquad
+\frac{(1+ (u-1)^2\bar{L}^2) \partial_{\bar{L}} u_0 + (u-1)^2 \bar{L} E\partial_E u_0}{(u_0 -1)^2 u_0 \nu_0}
\Big) .
   \end{aligned}
\end{align}
\end{subequations}

For the LRL proper time $\T$, given by expression \eqref{Tau}:
\begin{equation}\label{XtilTau}
  \X_{(\T)} = \P^t_{(\T)}\partial_t + \P^r_{(\T)}\partial_r + \P^\phi_{(\T)}\partial_\phi 
  +N^{-1}\T\X_{(N)} 
\end{equation}      
has components
\begin{subequations}\label{X.tilTau.components}
\begin{align}
& \begin{aligned}
    \P^t_{(\T)} = 
    \frac{2 M}{u}\Big(
    \int_{u_0}^{u} \frac{(E^2-u)\partial_E \nu + E \bar{L}\partial_{\bar{L}} \nu}{(y -1)^2 \nu^2} \, dy
    + \frac{(E^2-u) \partial_E u_0+ E \bar{L}\partial_{\bar{L}} u_0}{(u_0 -1)^2 \nu_0}
    \Big) , 
  \end{aligned}
\\
& \begin{aligned}
    \P^r_{(\T)} =
    2 M v \Big(
    \int_{u_0}^{u} \frac{E\partial_E \nu+ \bar{L}\partial_{\bar{L}} \nu}{(y -1)^2 \nu^2} \, dy
    + \frac{E\partial_E u_0 + \bar{L}\partial_{\bar{L}} u_0}{(u_0 -1)^2\nu_0}
    \Big), 
  \end{aligned}
\\
& \begin{aligned}
    \P^\phi_{(\T)} =
    \int_{u_0}^{u} & 
    \frac{(1 + (u -1)^2 \bar{L}^2)\partial_{\bar{L}} \nu + (u-1)^2\bar{L} E\partial_E \nu}{(y -1)^2\nu^2} \, dy
     \\&\quad
    + \frac{(1 + (u -1)^2\bar{L}^2)\partial_{\bar{L}} u_0 +(u-1)^2\bar{L} E \partial_E u_0}{(u_0 -1)^2\nu_0} .
     \end{aligned}
\end{align}
\end{subequations}

For $r_0$ chosen to be a horizon-crossing point or a centripetal point,
all of the preceding components $P^t$, $P^r$, $P^\phi$ are manifestly non-singular.
When $r_0$ is a turning point, note that 
$\nu$ will vanish at the endpoint $u=u_0$, and similarly $\nu_0$ will be zero,
which produces singularities.
However, these singularities can be shown to cancel,
so that the components $P^t$, $P^r$, $P^\phi$ remain non-singular. 

*** REMOVE SINGULAR TERMS FOR r0=TP

The physical meaning of these generators \eqref{XtilPhi}, \eqref{XtilT}, \eqref{XtilTau}
will emerge from how they act on the conserved quantities,
as shown in the next section. 
The associated groups of symmetry transformations
given by Proposition~\ref{prop:symm.group} 
will be derived in section~\ref{sec:transformations}.

\section{Commutators of LRL symmetries and action on conserved quantities}\label{sec:commutators.symmactions}

Three main results about the variational symmetries will be presented here.
First, the action of each symmetry on the conserved quantities
$L$, $E$, $\Phi$, $T$, $\T$ is obtained,
which leads to a physical interpretation of the symmetries. 
Second, the commutators of the symmetries are derived,
showing how the symmetries form a Poisson algebra. 
Third, the structure of the Poisson algebra will be presented
in terms of the variational symmetries arising from $L$, $E$, $\Phi$, $T$, $\T$.

\subsection{Symmetry actions}

A variational symmetry $\X_{(C_1)}$ acts on a locally conserved quantity $C_2$ 
by applying the prolonged symmetry operator $\pr\X_{(C_1)}$ to $C_2$ expressed as 
a function of $\lambda$, $t$, $r$, $\phi$, $\dot{t}$, $\dot{r}$, $\dot{\phi}$, 
and evaluating the result on the solution space $\solnsp$ of the equatorial geodesic equations.
As explained in section~\ref{sec:noether},
this yields a conserved quantity \eqref{prX.C}
whose dependence on $C_1$ and $C_2$ is antisymmetric \eqref{prX.C1.C2}. 
In particular, the action of $\pr\X_{(C)}$ on $C$ itself vanishes.

In addition, note that the action of the re-parameterization symmetries \eqref{X.re-param} 
on a conserved quantity automatically vanishes.
This is a consequence of 
\begin{equation}
  \pr\X_\triv C = f(\lambda,x,\dot{x}) \frac{dC}{d\tau}=0
\end{equation}
holding on $\solnsp$. 

Start with the variational symmetries \eqref{equator.X.L.E}
representing the Killing vectors \eqref{equator.KVs}.
Their action on the angular momentum $L$ and energy $E$
is given by 
\begin{equation}\label{L.E.actions.L.E}
  \pr\X_{(L)}E =- \pr\X_{(E)}L =0 . 
\end{equation}
This result is expected because the isometries generated by the Killing vectors
have a natural action on functions defined along any geodesic in spacetime,
which will vanish for any function that does not explicitly contain $t$ and $\phi$. 

The action of these two variational symmetries
on the three LRL conserved quantities $\Phi$, $T$, $\T$
is found to be
\begin{gather}
\pr\X_{(L)}\Phi =1, 
\quad
\pr\X_{(E)}\Phi =0,
\quad
\pr\X_{(L)}T =0,
\quad
\pr\X_{(E)}T =-1,
\label{L.E.actions.Phi.T}
\\
\pr\X_{(L)}\T =-L,
\quad
\pr\X_{(E)}\T =-E , 
\label{L.E.actions.Tau}
\end{gather}
by using chain rule combined with the previous symmetry actions \eqref{L.E.actions.L.E}.
Then the property \eqref{prX.C1.C2} gives 
\begin{gather}
\pr\X_{(\Phi)}L =-1,
\quad
\pr\X_{(\Phi)}E =0,
\quad
\pr\X_{(T)}L =0,
\quad
\pr\X_{(T)}E =1,
\label{Phi.T.actions.L.E}
\\
\pr\X_{(\T)}L =L,
\quad
\pr\X_{(\T)}E =E .
\label{Tau.actions.L.E}
\end{gather}

The result \eqref{Phi.T.actions.L.E} shows that 
$\Phi$ acts as a symmetry operator to shift $L$ and preserve $E$,
while $T$ acts to shift $E$ and preserve $L$.
This is similar to how the analogous LRL symmetries in Newtonian gravity
act on the conserved angular momentum and energy
in the plane of motion \cite{Anc.Mea.Pas,Ban.Itz}.
Likewise, the result \eqref{Tau.actions.L.E} indicates that $\T$ acts to scale
both $L$ and $E$ such that their ratio $L/E$ remains invariant.
This can be understood from the reciprocal actions \eqref{L.E.actions.Tau}
which arise because the symmetry operators \eqref{equator.X.L.E} 
generated by $L$ and $E$ 
contain an infinitesimal proper-time translation that acts on the $\tau$ term in $\T$. 

Similarly,
the action of the variational symmetries \eqref{XtilPhi}, \eqref{XtilT}, \eqref{XtilTau}
on the LRL conserved quantities themselves
consists of 
\begin{align}
& \pr\X_{(T)}\Phi = -\pr\X_{(\Phi)} T  = C_{T,\Phi},  
\label{T.Phi.actions}\\
& \pr\X_{(\T)} \Phi = -\pr\X_{(\Phi)} \T =  C_{\T,\Phi},
\label{Tau.Phi.actions}\\
& \pr\X_{(\T)} T = -\pr\X_{(T)} \T =C_{\T,T},
\label{Tau.T.actions}
\end{align}
where the righthand sides are conserved quantities
given by functions of $\bar{L}$ and $E$:
\begin{subequations}\label{C.LRL}
\begin{align}
C_{T,\Phi} & = \frac{\bar{L} (u_0-1)^2 u_0 \partial_E u_0 +E \partial_{\bar{L}} u_0}{(u_0 -1)^2 u_0\nu_0} , 
\\
C_{\T,\Phi} & =
\frac{(\bar{L}^2(u_0-1)^2 + 1) \partial_{\bar{L}} u_0 + \bar{L} E (u_0-1)^2 \partial_E u_0}{(u_0-1)^2 \nu_0} , 
\\
C_{\T,T} & =  \frac{2M ( (E^2 -u_0) \partial_E u_0 + \bar{L} E\partial_{\bar{L}} u_0 )}{(u_0-1)^2 u_0\nu_0} , 
\end{align}
\end{subequations}
with 
$|\nu_0|=\sqrt{(E^2 - u_0)\bar{L}^{-2} - u_0(u_0 -1)^2}$.
So far, $u_0$ is an arbitrary function of $\bar{L}$ and $E$.
When $r_0$ is chosen to be a dynamically distinguished radial position,
namely a turning point \eqref{tp.eqn}
or a centripetal point \eqref{cp.eqn}
or a horizon-crossing point,
then $u_0$ respectively satisfies 
$u_0^3 - 2 u_0^2 + (1+\bar{L}^{-2})u_0 - E^2\bar{L}^{-2} =0$
or 
$3 u_0^2 - 4 u_0 + 1 + \bar{L}^{-2} =0$
or
$u_0=0$.
In the case of a turning point or a horizon-crossing point, 
the conserved quantities \eqref{C.LRL} vanish,
\begin{equation}\label{C.LRL.tp.hp}
  C_{T,\Phi} = C_{\T,\Phi} = C_{\T,T} = 0 .
\end{equation}
*** NEED NON-SINGULAR FORM of Cs

In the case of a centripetal point, 
these quantities are found to be non-vanishing: 
\begin{subequations}\label{C.LRL.cp}
\begin{align}
  C_{T,\Phi} & = \frac{E(1-3u_0^*)}{\bar{L}(3u_0^* - 2)(u_0^* -1) u_0^* \nu_0^*} , 
\label{C.TPhi.LRL.cp}
\\
C_{\T,\Phi} & = \frac{(1-3u_0^*)(1 + \bar{L}^2(u_0^*-1)^2)}{\bar{L}(3u_0^* - 2) (u_0^*-1) \nu_0^*} , 
\label{C.PhiTau.LRL.cp}
\\
C_{\T,T} & = \frac{2M E(1 -3u_0^*)}{(3u_0^* -2)(u_0^*-1) u_0^* \nu_0^*} , 
\label{C.TTau.LRL.cp}
\end{align}
\end{subequations}
where 
$u_0^*= \tfrac{1}{3}(2 \pm \sqrt{1 -3\bar{L}^{-2}})$,
and 
$\nu_0^*=\tfrac{2}{3}\sqrt{2\bar{L}^{-2}(u_0-1)  - 2(3u_0+1) +9E^2}$.

\subsection{Commutators}

The symmetry actions \eqref{L.E.actions.L.E}--\eqref{Tau.Phi.actions}
are closely related to the commutators of the variational symmetries
through the property \eqref{C1.C2.commutator} explained in section~\ref{sec:noether}.
This property, however, has two caveats.
First, it holds only modulo the re-parameterization symmetries \eqref{X.re-param}. 
Second, it applies only to symmetries whose components $P^\mu$
are given by local functions of $\lambda$, $x^\nu$, $\dot{x}^\nu$,
and so $\X_{(\T)}$ is consequently excluded, 
since $\tau = \int N\,d\lambda =\int \sqrt{-\Lagr}\,d\lambda$
is a non-local function. 

To circumvent these difficulties,
the commutators can be computed directly. 
The vanishing commutators are found to be 
\begin{equation}\label{comm.0}
  [\pr\X_{(L)},\pr\X_{(E)}] =0,
  \quad
  [\pr\X_{(E)},\pr\X_{(\Phi)}] =0,
  \quad
  [\pr\X_{(L)},\pr\X_{(T)}] =0, 
\end{equation}
while the commutators that yield a re-parameterization symmetry 
consist of
\begin{equation}\label{comm.not0}
\begin{gathered}
  [\pr\X_{(L)},\pr\X_{(\Phi)}] =\pr\X_\shft,
  \quad
  [\pr\X_{(E)},\pr\X_{(T)}] =-\pr\X_\shft,
  \\
  [\pr\X_{(L)},\pr\X_{(\T)}] =-L\pr\X_\shft,
  \quad
 [\pr\X_{(E)},\pr\X_{(\T)}] = -E\pr\X_\shft ,
\end{gathered}
\end{equation}
where $\X_\shft$ is an infinitesimal shift \eqref{equator.shift} of proper time,
and $\pr\X_\shft = \frac{d}{d\tau}$. 
The remaining commutators are non-trivial: 
\begin{subequations}\label{comm.Cs}
\begin{gather}
[\pr\X_{(T)},\pr\X_{(\Phi)}]
= \pr\X_{(C_{T,\Phi})} ,
\label{comm.TPhi}
\\
[\pr\X_{(\T)},\pr\X_{(\Phi)}]
= -\pr\X_{(\Phi)} + \pr\X_{(C_{\T,\Phi})} +(C_{\T,T}/E) X_{(E)} ,
\label{comm.TauPhi}
\\
[\pr\X_{(\T)},\pr\X_{(T)}]
= -\pr\X_{(T)} + \pr\X_{(C_{\T,T})} +(C_{\T,T}/L) X_{(L)} .
\label{comm.TauT}
\end{gather}
\end{subequations}
In the case when $r_0$ is a turning point or a horizon-crossing point, 
the conserved quantities  $C_{T,\Phi}$, $C_{\T,\Phi}$, $C_{\T,T}$ vanish,
and hence the corresponding variational symmetries are zero. 
In the case when $r_0$ is a centripetal point,
$C_{T,\Phi}$, $C_{\T,\Phi}$, $C_{\T,T}$
are non-vanishing functions \eqref{C.LRL.cp} of $L$ and $E$,
whereby the variational symmetries
$\pr\X_{(C_{T,\Phi})}$, $\pr\X_{(C_{\T,\Phi})}$, $\pr\X_{(C_{\T,T})}$ vanish,
become linear combinations of the Killing vector symmetries
$\pr\X_{(L)}$ and $\pr\X_{(E)}$.

\subsection{Algebraic structure}

The set of conserved quantities $\{L,E,\Phi,T,\T\}$
has a rich algebraic structure stemming from
the symmetry actions \eqref{L.E.actions.L.E}--\eqref{Tau.Phi.actions}. 
Firstly, as shown by the results in section~\ref{sec:structure}, 
the function space of conserved quantities $\{L,E,\Phi,T,\T\}$
constitutes a Poisson algebra with the Poisson bracket
$\{C_1,C_2\} = \pr\X_{(C_1)}C_2=C_3$,
where $C_i=C_i(L,E,\Phi,T,\T)$ are smooth functions.
This Poisson algebra can be shown to be isomorphic to
the algebra of Hamiltonian Poisson brackets of $\{L,E,\Phi,T,\T\}$
in the canonical Hamiltonian form of the geodesic equations
coming from the geodesic Lagrangian \eqref{equator.Lagr}. 

Secondly, 
the variational symmetries arising from $L$, $E$, $\Phi$, $T$, $\T$
close into a finite-dimensional Lie algebra
only when $r_0$ is a turning point or a horizon point,
where $C_{T,\Phi}=C_{\T,\Phi}=C_{\T,T}=0$. 
In those two cases, this Lie algebra constitutes the Noether symmetry algebra of
the equatorial non-circular geodesics. 
In contrast, when $r_0$ is a centripetal point, 
the righthand sides of the commutators \eqref{comm.Cs}
contain nonlinear functions of $L$ and $E$.
Consequently, a closed Poisson subalgebra exists only 
on the linear space of conserved quantities $\{f(L,E),\Phi,T,\T\}$
where $f$ denotes an arbitrary smooth function. 
This Poisson subalgebra has a natural Lie algebra structure,
which can be viewed as the Noether symmetry algebra of
the equatorial non-circular geodesics in the case of a centripetal point. 

\begin{thm}\label{thm:liealgebra}
The structure of the Noether symmetry Lie algebra $\alg{g}$
arising from the locally conserved quantities $\{L,E,\Phi,T,\T\}$
depends on whether the LRL quantities $\Phi,T,\T$ 
are formulated using a turning point, a centripetal point, or a horizon-crossing point.
Let $\alg{f} = \spans(f(L,E)\pr\X_\shft)$ generated by (smooth) functions $f$,
which defines an infinite-dimensional abelian Lie algebra. 
\\
(i) In the case of a turning point or a horizon-crossing point,
the Lie algebra is a semi-direct sum of subalgebras
\begin{equation}
 \alg{g} = \alg{h} \ltimes(\Rnum^2 \oplus \alg{f})
\end{equation}
where $\alg{f}$ is an ideal; 
$\Rnum^2 = \spans(\pr\X_{(L)},\pr\X_{(E)})$ is an abelian subalgebra; 
$\alg{h} = \spans(\pr\X_{(\Phi)},\pr\X_{(\T)},\pr\X_{(T)})$ is a 3-dimensional subalgebra
with non-zero Lie brackets
\begin{equation}
  [\pr\X_{(\T)},\pr\X_{(\Phi)}] = -\pr\X_{(\Phi)},
  \quad
 [\pr\X_{(\T)},\pr\X_{(T)}] = -\pr\X_{(T)} ; 
\end{equation}
and $\alg{h}$ acts on $\Rnum^2$ via the Lie brackets \eqref{comm.not0}.
\\
(ii) In the case of a centripetal point,
the Lie algebra is a direct sum of vector spaces
\begin{equation}
  \alg{g} = \Rnum^3 \oplus \alg{k}\oplus\alg{f}
\end{equation}
where $\alg{f}$ is an ideal;
$\alg{k} = \spans(\pr\X_{(f(L,E))})$ generated by (smooth) functions $f$
is an infinite-dimensional abelian subalgebra;
$\Rnum^3 = \spans(\pr\X_{(\Phi)},\pr\X_{(\T)},\pr\X_{(T)})$ is a subspace
with the Lie brackets \eqref{comm.Cs}
and
\begin{subequations}
  \begin{gather}
    [\pr\X_{(\Phi)},\pr\X_{(f(L,E))}] = -f_L(L,E)\pr\X_\shft, 
    \quad
    [\pr\X_{(T)},\pr\X_{(f(L,E))}] = f_E(L,E) \pr\X_\shft, 
    \\
      [\pr\X_{(\T)},\pr\X_{(f(L,E))}] = \pr\X_{(g(L,E))} +g(L,E)\pr\X_\shft,
      \quad
      g= Lf_L + Ef_E
  \end{gather}
\end{subequations}
\end{thm}

A remark is that $\alg{h}$ has the structure $L_{3,4}$
listed in the well-known classification of low dimensional Lie algebras in Ref.\cite{SnoWin}.

\section{Equatorial symmetry transformations}\label{sec:transformations}

The variational symmetries arising from
the three LRL conserved quantities $\Phi$, $T$, $\T$
each generate a local Lie group of symmetry transformations 
on the equatorial geodesic variables 
$(\tau,t,r,\phi,\frac{dt}{d\tau},\frac{dr}{d\tau},\frac{d\phi}{d\tau})$
as shown in Proposition~\ref{prop:symm.group}. 
These dynamical symmetry transformations
will now be worked out explicitly. 
Instead of directly constructing the exponential series that
defines the symmetry transformation group \eqref{X.group}
for a given variational symmetry, which is quite complicated, 
there are two alternative approaches.

One method would be to express each symmetry transformation group
$\exp\big(\varepsilon\pr\X_{(C)}\big)$
as the integral curve of a vector field in the space of variables 
$(t,r,\phi,\frac{dt}{d\tau},\frac{dr}{d\tau},\frac{d\phi}{d\tau})$
and solve the system of differential equations that defines this curve.
A simpler method is to start from the conserved quantities
$L$, $E$, $\Phi$, $T$, $\T$ expressed in terms of
$t$, $r$, $\phi$, $\frac{dt}{d\tau}$, $\frac{dr}{d\tau}$, $\frac{d\phi}{d\tau}$, 
and then make use of how each symmetry acts on the conserved quantities,
which coincides with the action of the infinitesimal transformation group. 
This approach has been used successfully in recent work \cite{Anc.Gol.2026}
to obtain the symmetry transformation group
generated by the LRL vector in Newtonian gravity
in an explicit form in terms of the dynamical variables.

To carry out the latter method, 
it will be advantageous to work in the extended space of variables
$(\tau,t,r,\phi,\sigma,v,\omega)$,
with $\sigma =\frac{dt}{d\tau}$, $v=\frac{dr}{d\tau}$, $\omega=\frac{d\phi}{d\tau}$,
which turns out to carry the following gauge freedom.
A variational symmetry generator $\pr\X$
can be represented by an equivalent generator 
\begin{equation}\label{Y}
\pr\Y= \chi N^{-1} D_\lambda +\pr\X
\end{equation}
where 
\begin{equation}
  D_\lambda = \partial_\lambda + \dot{t}\partial_t + \dot{r}\partial_r + \dot{\phi}\partial_\phi
  + \ddot{t}\partial_{\dot{t}} + \ddot{r}\partial_{\dot{r}} + \ddot{\phi}\partial_{\dot{\phi}}
\end{equation}
is a total derivative operator with respect to the affine parameter $\lambda$,
with $(\ddot{t},\ddot{r},\ddot{\phi})$ satisfying 
the Euler-Lagrange equations of the Lagrangian \eqref{equator.Lagr}.
This extended generator \eqref{Y} acts in the same way as $\pr\X$
on all geodesics $(t(\lambda),r(\lambda),\phi(\lambda))$, 
and in particular has the same action as $\pr\X$ on all locally conserved quantities $C$,
since $\dot{C}=N \frac{d}{d\tau}C =0$. 
Hence,
\begin{equation}\label{Y.group}
  (\tau,t,r,\phi,\sigma,v,\omega) \to
  (\tau^\dagger,t^\dagger,r^\dagger,\phi^\dagger,\sigma^\dagger,v^\dagger,\omega^\dagger)
  =  \exp\big(\varepsilon\pr\Y\big)(\tau,t,r,\phi,\sigma,v,\omega) 
\end{equation}
yields a transformation group which is equivalent to the one generated by $\pr\X$,
where $\varepsilon$ is the group parameter. 
This equivalence contains the gauge freedom 
consisting of $\chi$ being an arbitrary function of $(\tau,t,r,\phi,\sigma,v,\omega)$.
(For more discussion, see the general results in \Ref{Anc.2026} for Lagrangian dynamical systems.)

The goal will now be to derive the explicit form of this transformation group \eqref{Y.group}
for the three variational symmetries
\eqref{XtilPhi}, \eqref{XtilT}, \eqref{XtilTau}
arising from the LRL conserved quantities $\Phi$, $T$, $\T$.
A useful gauge choice will be to have $r$ be invariant:
$\pr\Y r = N^{-1}\dot{r}\chi +\X r=0$, 
which determines
\begin{equation}
  \chi = - P^r/v 
\end{equation}
where $v= N^{-1}\dot{r}$ is the radial speed \eqref{equator.v}.
Consequently, the generator \eqref{Y} is given by 
\begin{equation}
  \begin{aligned}
    \pr\Y = &
      (-P^r/v) \partial_\lambda
  + (P^t - P^r \dot{t}/v) \partial_t
  +(P^\phi -  P^r \dot{\phi}/v)\partial_\phi
\\&\quad  
  + (\tfrac{d}{d\lambda}P^t - P^r \ddot{t}/v) \partial_{\dot{t}}
  +(\tfrac{d}{d\lambda}P^\phi -  P^r \ddot{\phi}/v)\partial_{\dot{\phi}}
  +(\tfrac{d}{d\lambda}P^r - P^r \ddot{r}/v) \partial_{\dot{r}} . 
  \end{aligned}
\end{equation}
The form of this generator looks complicated,
but only its properties that
it leaves $r$ invariant and acts the same as $\pr\X$ on conserved quantities
will be needed. 

To proceed,
use expressions \eqref{ener}--\eqref{anglmom} and \eqref{Phi}--\eqref{v}
to get the relations
\begin{subequations}\label{dynvars}
\begin{gather}
  \tau = \T +\int_{r_0}^{r} \frac{dy}{v(y)},
  \quad
   t = T +  E \int_{r_0}^{r} \frac{y\, dy}{(y-2M) v(y)},
   \quad
    \phi = \Phi + L \int_{r_0}^{r} \frac{dy}{y^2 v(y)},
 \\
    \sigma = (1 - 2M/r)^{-1} E,
    \quad
    v = \sgn(v)\sqrt{E^2-(1- 2M/r)(1+L^2/r^2)},
    \quad
    \omega = r^{-2} L.
\end{gather}
\end{subequations}
Then the action of $\pr\Y$ on each of these variables 
can be directly derived from the equivalent action of $\pr\X$
on $L$, $E$, $\Phi$, $T$, and $\T$,
by using $\pr\Y r =0$.

Hereafter, a bar notation will be used wherever convenient
to denote a quantity divided by $2M$.

\subsection{Equatorial symmetry transformations generated by LRL angle}\label{sec:Phi.transformations}

For the LRL angular quantity $\Phi$, 
the symmetry actions \eqref{Phi.T.actions.L.E} and \eqref{T.Phi.actions}--\eqref{Tau.Phi.actions}
involving $\X_{(\Phi)}$ give the infinitesimal transformation equations
\begin{gather}
  \partial_\varepsilon L^\dagger = -1,
  \quad
  \partial_\varepsilon E^\dagger = 0,
\label{Phi.flow.L.E}
\\
  \partial_\varepsilon T^\dagger = -C_{T,\Phi}{}^\dagger,
  \quad
  \partial_\varepsilon \T^\dagger = -C_{\T,\Phi}{}^\dagger . 
\label{Phi.flow.T.Tau}
\end{gather}
Also note $\partial_\varepsilon \Phi^\dagger = 0$
holds due to the general property \eqref{prX.C.C}, 
and thus
\begin{equation}\label{Phi.group.Phi}
  \Phi^\dagger = \Phi . 
\end{equation}
The first two flow equations \eqref{Phi.flow.L.E}
yield
\begin{equation}\label{Phi.group.L.E}
  L^\dagger = L -\varepsilon,
  \quad
  E^\dagger = E,  
\end{equation}
while the last two flow equations \eqref{Phi.flow.T.Tau} reduce to quadrature
because their righthand sides are functions only of $L$ and $E$.

In the case of a turning point or a horizon-crossing point, 
the functions $C_{T,\Phi}$ and $C_{\T,\Phi}$ vanish,
whence
\begin{equation}\label{Phi.group.T.Tau.tp.hp}
  T^\dagger = T, 
  \quad
  \T^\dagger = \T .
\end{equation}
In the case of a centripetal point, 
the functions $C_{T,\Phi}$ and $C_{\T,\Phi}$ are non-vanishing,
which yields
\begin{equation}\label{Phi.group.T.Tau.cp}
  T^\dagger = T +\int_{L}^{L^\dagger} C_{T,\Phi} \,dL, 
  \quad
  \T^\dagger = \T +\int_{L}^{L^\dagger} C_{\T,\Phi} \,dL, 
\end{equation}
via changing the integration variable from $\varepsilon$ to $L$
through equation \eqref{Phi.group.L.E},
where $C_{T,\Phi}$ and $C_{\T,\Phi}$ are given by
expressions \eqref{C.TPhi.LRL.cp}--\eqref{C.PhiTau.LRL.cp}. 

Then the expressions \eqref{dynvars} lead directly to the following result.

\begin{thm}\label{thm:Phi.group}
The symmetry transformation group \eqref{Y.group}, with parameter $\varepsilon$,
generated by the LRL angular quantity $\Phi$
is explicitly given by
\begin{subequations}
  \begin{gather}
    r^\dagger =r,
    \\
    \tau^\dagger = \T^\dagger +\int_{r_0^\dagger}^{r} \frac{dy}{v^\dagger(y)},
    \quad
    t^\dagger = T^\dagger +  E \int_{r_0^\dagger}^{r} \frac{y\, dy}{(y-2M) v^\dagger(y)},
    \quad
    \phi^\dagger = \Phi + (L-\varepsilon) \int_{r_0^\dagger}^{r} \frac{dy}{y^2 v^\dagger(y)},
 \\
    \sigma^\dagger = (1 - 2M/r)^{-1} E,
    \quad
    \omega^\dagger = r^{-2} (L-\varepsilon)
\end{gather}
with the transformations \eqref{Phi.group.L.E} for $L$ and $E$, 
where
\begin{equation}
  v^\dagger(r) = \sgn(v)\sqrt{E^2-(1- 2M/r)(1+(L-\varepsilon)^2/r^2)} . 
\end{equation}
\end{subequations}
In the case when $r_0$ is a horizon-crossing point or a turning point, 
$T$ and $\T$ are invariant \eqref{Phi.group.T.Tau.tp.hp},
whereas in the case when $r_0$ is a centripetal point,
they have the transformations \eqref{Phi.group.T.Tau.cp}. 
A horizon-crossing point is invariant,
while the transformed turning point $r_0^\dagger$ satisfies 
\begin{equation}\label{Phi.group.tp.r0}
  (E^2 -1) r_0^\dagger{}^3 +2M r_0^\dagger{}^2 -(L-\varepsilon)^2 (r_0^\dagger -2M)=0
\end{equation}
where $|\bar{L}-\bar{\varepsilon}|\geq 2$ with
$E^2<E_+^\dagger{}^2 = \tfrac{2}{3}+\tfrac{2}{27}(\bar{L}-\bar{\varepsilon})^2\big( 1 + \sqrt{(1-3/(\bar{L}-\bar{\varepsilon})^2}^3 \big)$,
and $|\bar{L}-\bar{\varepsilon}|< 2$ with $E^2<1$. 
The transformed centripetal point $r_0^\dagger$ satisfies 
\begin{equation}\label{Phi.group.cp.r0}
  M r_0^\dagger{}^2 -(L-\varepsilon)^2 (r_0^\dagger - 3 M)=0
\end{equation}
where $|\bar{L}-\bar{\varepsilon}|\geq \sqrt{3}$ with 
$E^2\geq E_-^\dagger{}^2 = \tfrac{2}{3}+\tfrac{2}{27}(\bar{L}-\bar{\varepsilon})^2\big( 1 - \sqrt{1-3/(\bar{L}-\bar{\varepsilon})^2}^3 \big)$.
\end{thm}

The parameter space $\{\varepsilon\}$ in these transformations \eqref{Phi.group.Phi}--\eqref{Phi.group.cp.r0}
is constrained in the case of a turning point and a centripetal point.
The constraints are obtained by solving the inequalities respectively
following equations \eqref{Phi.group.tp.r0} and \eqref{Phi.group.cp.r0}.
(Details are shown in the Appendix.)
Let
\begin{equation}
  1/\bar{L}_\pm = \sqrt{ \tfrac{1}{2} - \tfrac{3}{2}(\tfrac{3}{2} E^2 -1)^2  \pm\tfrac{1}{8}E\sqrt{(9E^2-8)^3} },
\end{equation}  
which inverts the relation \eqref{E_pm}, 
namely, $|\bar{L}|=\bar{L}_\pm$ holds if and only if $E=E_\pm$.

For a centripetal point:
the inequalities yield
\begin{subequations}
\begin{align}
    & 3 < |\bar{L} - \bar{\varepsilon}|^2 < \bar{L}_-^2,
    \quad
    \tfrac{8}{9}< E^2<1 ;
    \\
    &  3 < |\bar{L} - \bar{\varepsilon}|^2,
    \quad
    1\leq E^2 . 
\end{align}
\end{subequations}
For a turning point:
the inequalities reduce to 
\begin{equation}
  \bar{L}_+^2 < |\bar{L}-\bar{\varepsilon}|^2, 
  \quad
  E^2 > 1 . 
\end{equation}

In both cases,
the set of constrained transformations
in a sufficiently small neighbourhood of the identity, $\varepsilon=0$,
comprise a one-dimensional local Lie group.
The physical meaning of this symmetry group 
comes from its action on $L$ and $E$ in the equatorial geodesics,
namely $E$ is invariant while $L$ gets shifted by $-\varepsilon$.
A shift of $L$ produces a change in the shape of the effective potential
which determines the different types of allowed orbits
(cf Tables~\ref{orbits-bounded} and ~\ref{orbits-unbounded}). 
If $E^2 >\tfrac{8}{9}$ then for a range of $|L|$ there exists more than one type of orbit,
and the symmetry transformations can map one type into another type:
for instance an elliptic-like orbit can be transformed into an asymptotic circular orbit. 
However, elliptic-like, parabolic-like, and hyperbolic-like orbits
cannot be transformed into each other, because they exist for different ranges of $E$,
similarly to the situation in Newtonian gravity. 
In contrast, if $E^2<\tfrac{8}{9}$ then for all $|L|\neq0$ there is only one type of orbit.

\subsection{Equatorial symmetry transformations generated by LRL time}\label{sec:T.transformations}

For the LRL temporal quantity $T$, 
the symmetry actions \eqref{Phi.T.actions.L.E}, \eqref{T.Phi.actions}, \eqref{Tau.T.actions} 
involving $\X_{(T)}$ give the infinitesimal transformation equations
\begin{gather}
  \partial_\varepsilon L^\dagger = 0,
  \quad
  \partial_\varepsilon E^\dagger = 1,
\label{T.flow.L.E}
\\
  \partial_\varepsilon \Phi^\dagger = C_{T,\Phi}{}^\dagger,
  \quad
  \partial_\varepsilon \T^\dagger = -C_{\T,T}{}^\dagger.
\label{T.flow.Phi.Tau}
\end{gather}
In addition, 
$\partial_\varepsilon T^\dagger = 0$ 
holds due to the general property \eqref{prX.C.C}, 
so that 
\begin{equation}\label{T.group.T}
  T^\dagger = T.
\end{equation}
The first two flow equations \eqref{T.flow.L.E}
yield
\begin{equation}\label{T.group.L.E}
  L^\dagger = L, 
  \quad
  E^\dagger = E + \varepsilon,  
\end{equation}
while the last two flow equations \eqref{T.flow.Phi.Tau} reduce to quadrature
because their righthand sides are functions only of $L$ and $E$.

In the case of a turning point or a horizon-crossing point, 
the functions $C_{T,\Phi}$ and $C_{\T,T}$ vanish,
and thus
\begin{equation}\label{T.group.Phi.Tau.tp.hp}
  \Phi^\dagger = \Phi, 
  \quad
  \T^\dagger = \T .
\end{equation}
In the case of a centripetal point, 
the functions $C_{T,\Phi}$ and $C_{\T,T}$ are non-vanishing,
which yields
\begin{equation}\label{T.group.Phi.Tau.cp}
  \Phi^\dagger = \Phi +\int_{E}^{E^\dagger} C_{T,\Phi} \,dE
  \quad
  \T^\dagger = \T -\int_{E}^{E^\dagger} C_{\T,T} \,dE,
\end{equation}
via changing the integration variable from $\varepsilon$ to $E$
through equation \eqref{T.group.L.E},
where $C_{T,\Phi}$ and $C_{\T,T}$ are given by
expressions \eqref{C.TPhi.LRL.cp} and \eqref{C.TTau.LRL.cp}. 

Then the expressions \eqref{dynvars} give the following result.

\begin{thm}\label{thm:T.group}
The symmetry transformation group \eqref{Y.group}, with parameter $\varepsilon$,
generated by the LRL temporal quantity $T$
is explicitly given by
\begin{subequations}
  \begin{gather}
    r^\dagger =r,
    \\
    \tau^\dagger = \T^\dagger +\int_{r_0^\dagger}^{r} \frac{dy}{v^\dagger(y)},
    \quad
    t^\dagger = T +  (E +\varepsilon)\int_{r_0^\dagger}^{r} \frac{y\, dy}{(y-2M) v^\dagger(y)},
    \quad
    \phi^\dagger = \Phi + L \int_{r_0^\dagger}^{r} \frac{dy}{y^2 v^\dagger(y)},
 \\
    \sigma^\dagger = (1 - 2M/r)^{-1} (E+\varepsilon),
    \quad
    \omega^\dagger = r^{-2} L, 
\end{gather}
with the transformations \eqref{T.group.L.E} for $L$ and $E$, 
where
\begin{equation}
  v^\dagger(r) = \sgn(v)\sqrt{(E+\varepsilon)^2-(1- 2M/r)(1+L^2/r^2)} . 
\end{equation}
\end{subequations}
In the cases when $r_0$ is a horizon-crossing point or a turning point, 
$\Phi$ and $\T$ are invariant \eqref{T.group.Phi.Tau.tp.hp},
along with the horizon-crossing point,
while the transformed turning point $r_0^\dagger$ satisfies 
\begin{equation}\label{T.group.tp.r0}
  ((E+\varepsilon)^2 - 1) r_0^\dagger{}^3 + 2M r_0^\dagger{}^2 - L^2(r_0^\dagger - 2M) =0
\end{equation}
where 
$E+\varepsilon \leq 
\begin{cases} E_+, & |\bar{L}|>2
\\ 1, & |\bar{L}|\leq 2 \end{cases}$. 
In the case when $r_0$ is a centripetal point,
$\Phi$ and $\T$ have the transformations \eqref{T.group.Phi.Tau.cp},
with $r_0$ being invariant and $|\bar{L}|\geq \sqrt{3}$, 
where  $E+\varepsilon \geq E_-$.
\end{thm}

The parameter space $\{\varepsilon\}$ in these transformations \eqref{T.group.T}--\eqref{T.group.tp.r0}
is constrained in the case of a turning point and a centripetal point.
For a centripetal point: the inequality involving $E$ implies
\begin{equation}
  \varepsilon > E_- - E
\end{equation}
with $|\bar{L}|\geq \sqrt{3}$ and $E>E_-$. 
For a turning point: the inequality involving $E$ yields two branches.
The first branch consists of the interval 
\begin{equation}
  -E < \varepsilon < 1- E,
  \quad
  |\bar{L}| \leq 2
\end{equation}
with $E<1$.
The second branch is given by
\begin{equation}
  -E < \varepsilon < E_+ - E,
  \quad
  |\bar{L}| >2
\end{equation}
with $E<E_+$.

In both cases,
the set of constrained transformations
in a sufficiently small neighbourhood of the identity, $\varepsilon=0$,
comprise a one-dimensional local Lie group.
The physical meaning of this symmetry transformation group
comes from how it acts on $L$ and $E$ in the equatorial geodesics,
namely $L$ is invariant while $E$ gets shifted by $\varepsilon$.
Consequently, there is no change in the shape of the effective potential
under the transformations.
For a range of $E$ there exists more than one type of orbit
(cf Tables~\ref{orbits-bounded} and ~\ref{orbits-unbounded}),
and the transformations can map these different types into each other. 
For instance, if $|\bar{L}| >2$ then an elliptic-like orbit can be transformed into
a parabolic-like orbit or a hyperbolic-like orbit. 
This is similar to the situation in Newtonian gravity. 
In contrast, if $|\bar{L}| <\sqrt{3}$ then for all $E>0$ there is only one type of orbit.

\subsection{Equatorial symmetry transformations generated by LRL proper time}\label{sec:Tau.transformations}

For the LRL proper time quantity $\T$, 
the symmetry actions \eqref{Tau.actions.L.E} and \eqref{Tau.Phi.actions}--\eqref{Tau.T.actions}
involving $\X_{(\T)}$
give the infinitesimal transformation equations
\begin{gather}
  \partial_\varepsilon L^\dagger = L^\dagger,
  \quad
  \partial_\varepsilon E^\dagger = E^\dagger
\label{Tau.flow.L.E}
\\
  \partial_\varepsilon \Phi^\dagger = C_{\T,\Phi}{}^\dagger,
  \quad
  \partial_\varepsilon T^\dagger = C_{\T,T}{}^\dagger . 
\label{Tau.flow.Phi.T}
\end{gather}

The first two flow equations \eqref{Tau.flow.L.E}
give 
\begin{equation}\label{Tau.group.L.E}
  L^\dagger = L e^\varepsilon,
  \quad
  E^\dagger = E e^\varepsilon .
\end{equation}
The last two flow equations \eqref{Tau.flow.Phi.T} reduce to quadrature
because their righthand sides do not contain $\Phi^\dagger$ and $T^\dagger$.

In the case of a turning point or a horizon-crossing point, 
the functions $C_{\T,\Phi}$ and $C_{\T,T}$ vanish,
and thus
\begin{equation}\label{Tau.group.Phi.T.tp.hp}
  \Phi^\dagger = \Phi, 
  \quad
  T^\dagger = T.
\end{equation}
In the case of a centripetal point, 
the functions $C_{\T,\Phi}$ and $C_{\T,T}$ are non-vanishing,
which yields
\begin{subequations}\label{Tau.group.Phi.T.cp}
  \begin{align}
    \Phi^\dagger & = \Phi +\int_{\sqrt{|LE|}}^{\sqrt{|LE|}^\dagger} C_{\T,\Phi}(\sgn(L) y \sqrt{|L|/|E|},\sgn(E) y \sqrt{|E|/|L|}) \,\frac{dy}{y},
    \\
    T^\dagger & = T +\int_{\sqrt{|LE|}}^{\sqrt{|LE|}^\dagger} C_{\T,T}(\sgn(L) y \sqrt{|L|/|E|},\sgn(E) y \sqrt{|E|/|L|}) \,\frac{dy}{y}
  \end{align}
\end{subequations}
where $C_{\T,\Phi}$ and $C_{\T,T}$ are given by
expressions \eqref{C.PhiTau.LRL.cp} and \eqref{C.TTau.LRL.cp}.

Finally, the general property \eqref{prX.C.C} shows that 
$\partial_\varepsilon \T^\dagger = 0$, 
whence 
\begin{equation}\label{Tau.group.Tau}
  \T^\dagger = \T .
\end{equation}
Now the expressions \eqref{dynvars} yield the following result.

\begin{thm}\label{thm:Tau.group}
The symmetry transformation group \eqref{Y.group}, with parameter $\varepsilon$,
generated by the LRL proper-time quantity $\T$
is explicitly given by
\begin{subequations}
  \begin{gather}
    r^\dagger =r,
    \\
    \tau^\dagger = \T +\int_{r_0^\dagger}^{r} \frac{dy}{v^\dagger(y)},
    \quad
    t^\dagger = T^\dagger + E^\dagger \int_{r_0^\dagger}^{r} \frac{y\, dy}{(y-2M) v^\dagger(y)},
    \quad
    \phi^\dagger = \Phi + L^\dagger \int_{r_0^\dagger}^{r} \frac{dy}{y^2 v^\dagger(y)},
 \\
    \sigma^\dagger = (1 - 2M/r)^{-1} E^\dagger, 
    \quad
    \omega^\dagger = r^{-2} L^\dagger, 
\end{gather}
with the transformations \eqref{Tau.group.L.E} for $L$ and $E$, 
where
\begin{equation}
  v^\dagger(r) = \sgn(v)\sqrt{E^\dagger{}^2 -(1- 2M/r)(1+(L^\dagger/r)^2)} .
\end{equation}
\end{subequations}
In the cases when $r_0$ is a turning point or a horizon-crossing point,
$\Phi$ and $T$ are invariant \eqref{Tau.group.Phi.T.tp.hp}, 
with $r_0$ being invariant for the latter case, 
while for the former case, the transformed turning point $r_0^\dagger$ satisfies 
\begin{equation}\label{Tau.group.tp.r0}
  (E^\dagger{}^2 - 1) r_0^\dagger{}^3 + 2M r_0^\dagger{}^2 - L^\dagger{}^2 (r_0^\dagger - 2M) =0 ,
\end{equation}
where
$E^2 \leq 
\begin{cases} E_+^\dagger{}^2 e^{-2\varepsilon}, & |\bar{L}| > 2e^{-\varepsilon}
  \\ e^{-2\varepsilon}  , & |\bar{L}|\leq 2 e^{-\varepsilon}\end{cases}$. 
In the case when $r_0$ is a centripetal point,
$\Phi$ and $\T$ have the transformations \eqref{T.group.Phi.Tau.cp},
and 
\begin{equation}\label{Tau.group.cp.r0}
  M r_0^\dagger{}^2 -(LN/N^\dagger)^2(r_0^\dagger - 3 M)=0
\end{equation}
determines the transformed centripetal point $r_0^\dagger$,
where
$E^2 \geq E_-^\dagger{}^2 e^{-2\varepsilon}$,
$|\bar{L}|\geq \sqrt{3}e^{-\varepsilon}$.
Here,
$E_\pm^\dagger{}^2 
= \tfrac{2}{3}+\tfrac{2}{27}\bar{L}^\dagger{}^2 \big( 1 \pm \sqrt{1-3/\bar{L}^\dagger{}^2}^3 \big)$.
\end{thm}

The parameter space $\{\varepsilon\}$ in these transformations \eqref{Tau.group.Tau}--\eqref{Tau.group.cp.r0}
is constrained in the case of a turning point and a centripetal point,
through solving the inequalities respectively
following equations \eqref{Tau.group.tp.r0} and \eqref{Tau.group.cp.r0}.
(Details are shown in the Appendix.)
Let $q>0$ denote the positive roots of the cubic 
$(1/b^2 - 1 - 9q)^2 = (1-3q)^3$, 
where
\begin{equation}\label{b}
  b^2=\tfrac{2}{27}(\bar{L}/E)^2
\end{equation}
is a scaled impact factor squared. 
Then the solution of $E^2/\bar{L}^2 = E_\pm^\dagger{}^2/\bar{L}^\dagger{}^2$
for $\varepsilon$ is given by
\begin{equation}\label{q}
  e^{-\varepsilon}= \sqrt{q}|\bar{L}| . 
\end{equation}

For a centripetal point: the inequalities reduce to 
\begin{equation}
  e^{-2\varepsilon} <
  \begin{cases}
    q_{\max} \bar{L}^2, & b^{-2} < 4 \\
    \tfrac{1}{3} \bar{L}^2, & 4 \leq b^{-2}
  \end{cases} .
\end{equation}
For a turning point: the inequalities yield 
\begin{equation}
  e^{-2\varepsilon} >
  \begin{cases}
    q_{\min}\bar{L}^2 , &   2 < b^{-2} \leq \tfrac{27}{8}
    \\
    \tfrac{2}{27} b^{-2}\bar{L}^2, & b^{-2} > \tfrac{27}{8}
   \end{cases} .
\end{equation}

In both cases,
the set of constrained transformations 
in a sufficiently small neighbourhood of the identity, $\varepsilon=0$,
comprise a one-dimensional local Lie group.
The physical meaning of this symmetry transformation group
involves a change in the shape of the effective potential as well as in the type of orbit, 
such that both $L$ and $E$ get scaled by $e^\varepsilon$,
while $E/L$ is invariant.

\subsection{Equatorial Noether symmetry group}
Compositions of the one-dimensional symmetry transformation groups
in the preceding Theorems~\ref{thm:Phi.group} to~\ref{thm:Tau.group}
together with the isometry groups generated by
the equatorial Killing vectors \eqref{equator.KVs}
generates the Noether symmetry group $\G$
of the equatorial non-circular geodesics,
modulo the re-parameterization symmetries $f(L,E)\pr\X_\shft$
(involving smooth functions $f$). 
The transformation groups respectively arising from the LRL quantities $\Phi$, $T$, $\T$
are subgroups of $\G$ having the structure of one-dimensional local Lie groups
$\G_\Phi\simeq \Rnum$, $\G_T\simeq\Rnum$, $\G_\T\simeq\Rnum$
due to the constraints on their parameters,
whereas the symmetry transformation groups
which arise from $L$ and $E$ are one-dimensional global Lie groups
$\G_L = U(1)$ and $\G_E= \Rnum$.
The structure of $\G$ as a local Lie group 
is determined by its Lie algebra given in Theorem~\ref{thm:liealgebra}.

When the LRL quantities are formulated using a turning point or a horizon-crossing point,
the Noether symmetry group $\G$ modulo re-parameterization symmetries 
is finite dimensional.
In contrast, when a centripetal point is used to formulate the LRL quantities,
the Noether symmetry group $\G$ is essentially infinite dimensional,
because the composition of $\G_\Phi$, $\G_T$, $\G_\T$ produces
elements in the group generated by symmetries of the form $\pr\X_{(f(L,E))}$
(involving smooth functions $f$).

\section{Non-equatorial geodesics}\label{sec:nonequatorial}

The main results on conserved quantities and variational symmetries
for equatorial timelike geodesics
will be extended here to non-equatorial timelike geodesics. 
A simple group-theoretical derivation is explained
which utilizes the spherical symmetry of 
the timelike geodesic equations \eqref{gen.geodesic.eqns}
in Schwarzschild spacetime.

\subsection{A complete set of conserved quantities}\label{subsec:cons.quantities}

The Schwarzschild metric possesses three rotational Killing vectors \cite{Mis.Tho.Whe,Wal} 
\begin{equation}\label{rotation.KVs}
  \xi_{(x)} = -\sin\phi \partial_\theta -\cos\phi \cot\theta \partial_\phi,
  \quad
  \xi_{(y)} = \cos\phi \partial_\theta -\sin\phi \cot\theta \partial_\phi,
  \quad
  \xi_{(z)} = \partial_\phi
\end{equation}
respectively representing infinitesimal rotations about the $x$, $y$, $z$ spatial directions
in the spacelike hypersurfaces $\Sigma$
orthogonal to the temporal Killing vector $\partial_t$.
Each Killing vector yields a conserved quantity of the geodesic equations \eqref{gen.geodesic.eqns}:
\begin{equation}\label{Lx.Ly.Lz}
  L^x = g(\xi_{(x)},\u),
  \quad
  L^y = g(\xi_{(y)},\u),
  \quad
  L^z = g(\xi_{(z)},\u)
\end{equation}
where $\u$ is the 4-velocity \eqref{4vel} of a geodesic. 
These three quantities describe angular momenta.
The Killing vector $\partial_t$ also produces a conserved quantity,
which is the energy
\begin{equation}\label{E}
  E = -g(\partial_t,\u) . 
\end{equation}

The timelike geodesic equations \eqref{gen.geodesic.eqns}
possess additional conserved quantities.
Similarly to the discussion in section~\ref{sec:conserved.quantities},
these equations describe a dynamical system for the spacetime motion of
a test particle on a geodesic.
As the dynamics has four degrees of freedom,
there will exist eight locally conserved quantities
$C(\tau,t,r,\phi,\theta,\frac{dt}{d\tau},\frac{dr}{d\tau},\frac{d\phi}{d\tau},\frac{d\theta}{d\tau})$
satisfying 
\begin{equation}\label{C}
  \frac{dC}{d\tau} =0
\end{equation}
piecewise in $\tau$ for all timelike geodesics,
where $\tau$ is proper time.
Apart from the three angular momenta $L^x$, $L^y$, $L^z$, and the energy $E$, 
a fifth conserved quantity is the norm of the 4-velocity
\begin{equation}\label{4vel.norm}
  -1 = g(\u,\u) . 
\end{equation}

Three more conserved quantities consist of
an LRL angle $\Phi$ in the plane of motion,
an LRL coordinate-time $T$ and proper time $\T$,
which are counterparts of the LRL quantities for non-radial equatorial geodesics.
These can be derived from the conservation equation \eqref{C}
by the same method used in the equatorial case.
Alternatively,
the following more direct, shorter method can be used.

The Killing vectors \eqref{rotation.KVs} 
generate an isometry group, $SO(3)$, of rotations acting on $\Sigma$. 
Under a rotation, the angular momenta \eqref{Lx.Ly.Lz} will transform
as components of a 3-dimensional vector.
Thus, it is natural to introduce an auxiliary vector space $\Rnum^3$
with orthonormal basis vectors $(1,0,0)$, $(0,1,0)$, $(0,0,1)$ 
associated to the $x$, $y$, $z$ rotation axes of the Killing vectors in $\Sigma$.
Then the angular momenta can be viewed as representing a vector quantity 
$(L^x,L^y,L^z)$ in $\Rnum^3$.
Note that this auxiliary vector space is not part of the Schwarzschild spacetime,
but rather it serves as a representation space for the $SO(3)$ rotation group.

The angular quantity $\Phi$ describes a direction angle in the plane of motion.
This direction corresponds to a unit vector $(\A^x,\A^y,\A^z)$
in the auxiliary vector space $\Rnum^3$, 
with $(\A^x)^2+(\A^y)^2+(\A^z)^2=1$. 
Explicit expressions for its components arise
from considering a $SO(3)$ rotation under which the plane of motion is carried to
the equatorial plane $\theta=\frac{1}{2}\pi$.
The requisite rotation can be taken as the product of
a polar rotation and an azimuthal rotation 
applied to the angular momentum vector: 
first use an azimuthal rotation $R_{(\phi)}\in SO(3)$ that carries
$(L^x,L^y,L^z)$ to $(\tilde L^x,0,L^z)$; 
then use a polar rotation $R_{(\theta)} \in SO(3)$ that carries
$(\tilde L^x,0,L^z)$ to $(0,0,L)$.
These rotations are given by the matrices
\begin{equation}\label{azimuth.rotation}
  R_{(\phi^*)} =
  \begin{pmatrix}
    \cos\phi^* & -\sin\phi^* & 0 \\
    \sin\phi^* & \cos\phi^* & 0 \\
    0 & 0 & 1
  \end{pmatrix},
  \quad
  \phi^* = -\arctan(L^y/L^x)
\end{equation}
and
\begin{equation}\label{polar.rotation}
  R_{(\theta^*)} =
  \begin{pmatrix}
    \cos\theta^* & 0 & \sin\theta^* \\
    0 & 1 & 0 \\
    -\sin\theta^* & 0 & \cos\theta^*
  \end{pmatrix},
  \quad
  \theta^* =-\sgn(L^x)\arctan\big(\sqrt{(L^x)^2+(L^y)^2}/L^z\big) . 
\end{equation}
The resulting angular momentum vector $(0,0,L)$ is given by
\begin{equation}\label{L}
  L = \sgn(L^z)\sqrt{(L^x)^2 + (L^y)^2 + (L^z)^2}
  = L^z\sqrt{1+ (L^x/L^z)^2 + (L^y/L^z)^2} .
\end{equation}
This coincides with the scalar angular momentum \eqref{anglmom}
in the case of equatorial geodesics.
Now apply the inverse rotations to the equatorial LRL direction angle
represented as the unit vector $(\cos\Phi,\sin\Phi,0)$ in $\Rnum^3$, 
where $\Phi$ is the LRL angle \eqref{Phi}. 
This gives
\begin{equation}\label{A.Phi}
(\A^x,\A^y,\A^z)^\t = R_{(\theta^*)}^{-1}R_{(\phi^*)}^{-1} (\cos\Phi,\sin\Phi,0)^\t
\end{equation}
with ``$\t$'' denoting the transpose,
where the components are given by 
\begin{subequations}\label{Ax.Ay.Az}
  \begin{align}
     \A^x & =
     \tfrac{|L^x|\,|L^z|}{\sqrt{(L^x)^2 + (L^y)^2}\sqrt{(L^x)^2 + (L^y)^2 + (L^z)^2}}  \cos(\Phi)
  - \tfrac{\sgn(L^x) L^y}{\sqrt{(L^x)^2 + (L^y)^2}} \sin(\Phi), 
  \\
  \A^y & =
  \tfrac{\sgn(L^x) L^y |L^z|}{\sqrt{(L^x)^2 + (L^y)^2}\sqrt{(L^x)^2 + (L^y)^2 + (L^z)^2}} \cos(\Phi)
  + \tfrac{|L^x|}{\sqrt{(L^x)^2 + (L^y)^2}} \sin(\Phi) , 
  \\
  \A^z & =
  -\tfrac{\sgn(L^x) \sgn(L^z)\sqrt{(L^x)^2 + (L^y)^2}}{\sqrt{(L^x)^2 + (L^y)^2 + (L^z)^2}} \cos(\Phi)  . 
\end{align}
\end{subequations}

After these rotations,
the expression for $\Phi$ gets changed by
replacing the angle $\phi$ in the equatorial plane 
by a corresponding angle $\alpha$ in the plane of motion.
This angle is determined from the dynamical condition
\begin{equation}
  N^{-1} r^2 \dot{\alpha} = L , 
\end{equation}
which yields 
\begin{equation}\label{alpha}
  \alpha = \arctan\Big( \frac{((n^x,n^y,n^z)\times (\sin\theta\cos\phi,\sin\theta\sin\phi,\cos\theta))\cdot (L^x,L^y,L^z)}{(n^x,n^y,n^z)\cdot (\sin\theta\cos\phi,\sin\theta\sin\phi,\cos\theta)L} \Big)
\end{equation}
up to addition of an arbitrary constant of motion,
where $(n^x,n^y,n^z)$ is a constant unit vector
orthogonal to $(L^x,L^y,L^z)$,
and 
$(\sin\theta\cos\phi,\sin\theta\sin\phi,\cos\theta)$ is the position vector on the celestial sphere.  
Geometrically,
$\alpha$ represents an angle in the invariable plane containing the geodesic,
with $\alpha=0$ being a radial line whose angular coordinates are given by
$(\sin\theta\cos\phi,\sin\theta\sin\phi,\cos\theta) = (n^x,n^y,n^z)$. 
In the terminology of orbital motion (see e.g. \Ref{Cho}), 
$(n^x,n^y,n^z)$ is analogous to the ``position of the ascending node''
while $\alpha$ is analogous to the ``argument of latitude''. 

Hence, the LRL angle for non-equatorial timelike geodesics is given by 
\begin{equation}\label{gen.Phi}
  \Phi = \alpha - \sgn(\tfrac{dr}{d\tau}) L \int_{r_0}^{r} \frac{dy}{y^2|v(y,L,E)|}
\end{equation}
where $E$ is the energy \eqref{E}, $L$ is the scalar angular momentum \eqref{L},
and 
\begin{equation}\label{absv}
|v(r,L,E)| = \sqrt{E^2-(1- 2M/r)(1+L^2/r^2)} . 
\end{equation}

Each component of the LRL direction angle $(\A^x,\A^y,\A^z)$
is a locally conserved quantity for timelike geodesics,
and therefore $\Phi$ is also locally conserved. 
However, $(\A^x,\A^y,\A^z)$ satisfies two relations
\begin{equation}\label{A.rels}
  (\A^x)^2+(\A^y)^2+(\A^z)^2=1,
  \quad
  L^x \A^x+L^y \A^y + L^z \A^z =0 , 
\end{equation}
and so $\Phi$ is the only part of $(\A^x,\A^y,\A^z)$ that is functionally independent of
$(L^x,L^y,L^z)$ and $E$.
Finally, 
the two LRL temporal quantities
\begin{equation}\label{gen.T}
  T = t - \sgn(\tfrac{dr}{d\tau}) E \int_{r_0}^{r} \frac{r\, dr}{(r-2M) |v(r,L,E)|}
\end{equation}
and
\begin{equation}\label{gen.Tau}
\T = \tau - \sgn(\tfrac{dr}{d\tau}) \int_{r_0}^{r} \frac{dr}{|v(r,L,E)|}
\end{equation}
are also locally conserved for non-equatorial timelike geodesics,
with $E$ and $L$ again given by expressions \eqref{E} and \eqref{L}. 

In the LRL quantities, 
$r_0$ is taken to be a turning point, a centripetal point, or a horizon-crossing point,
similarly to the equatorial case.
For non-equatorial geodesics,
there is no change in the radial speed equation \eqref{v},
and thus these dynamically distinguished radial points are determined by
the respective equations \eqref{tp.eqn}, \eqref{cp.eqn}, and $r_0=2M$. 

\begin{thm}\label{thm:cons.quantities}
The seven quantities $E$, $L^x$, $L^y$, $L^z$, $\Phi$, $T$, $\T$,
together with the 4-velocity norm \eqref{4vel.norm},
comprise a complete set of functionally-independent locally conserved quantities
for the non-equatorial timelike geodesic equations \eqref{gen.geodesic.eqns}.
\end{thm}

In particular,
these conserved quantities do not depend on initial values for the geodesics.

\subsection{Variational symmetries}\label{subsec:symms}

The Noether correspondence \eqref{P.C}
in Proposition~\ref{prop:noether.reverse}
shows that a variational symmetry of the Lagrangian for timelike geodesics
arises from each locally conserved quantity. 
Energy \eqref{E} and angular momenta \eqref{Lx.Ly.Lz} respectively correspond to
\begin{equation}\label{gen.X.E}
  \X_{(E)} =   -\partial_t + E \X_\shft
\end{equation}
and
\begin{equation}\label{gen.X.Lx.Ly.Lz}
  \X_{(L^i)} =   \xi_{(i)} + L^i \X_\shft, 
  \quad
  i = x,y,z
\end{equation}
where
\begin{equation}\label{gen.X.N}
  \X_\shft = \frac{dt}{d\tau}\partial_t + \frac{dr}{d\tau}\partial_r +\frac{d\phi}{d\tau}\partial_\phi +\frac{d\theta}{d\tau}\partial_\theta
\end{equation}
generates infinitesimal proper-time translations.
Modulo these translations,
the variational symmetry operators \eqref{gen.X.E}--\eqref{gen.X.Lx.Ly.Lz}
coincide with the form of the Killing vectors. 
It is important to remember, however, that  
the variational symmetries are defined to act on the variables
$(t,r,\phi,\theta,\dot{t},\dot{r},\dot{\phi},\dot{\theta})$ 
in the geodesic Lagrangian,
whereas the Killing vectors act in the spacetime.

The variational symmetries $\X_{(\Phi)}$, $\X_{(T)}$, $\X_{(\T)}$
arising from the three LRL quantities
have a similar form to the ones \eqref{XtilPhi}, \eqref{XtilT}, \eqref{XtilTau}
found in the equatorial case.
Since the $SO(3)$ rotations \eqref{azimuth.rotation}--\eqref{polar.rotation}
leave invariant the geodesic variables $t$, $r$, $\dot{t}$, $\dot{r}$,
this implies that both components $P^t_{(C)}$ and $P^r_{(C)}$ are unchanged,
while the components $P^\phi_{(C)}$ and $P^\theta_{(C)}$ get transformed as follows.
For each conserved quantity $C$ other than $\Phi$:
$P^\phi_{(C)} \to \frac{\partial L}{\partial \dot{\phi}} P^\phi_{(C)}$
and
$P^\theta_{(C)} \to \frac{\partial L}{\partial \dot{\theta}} P^\phi_{(C)}$.
For $C=\Phi$, 
$P^\phi_{(C)} \to \frac{\partial L}{\partial \dot{\phi}} P^\phi_{(C)} + P^\phi_{(\alpha)}$
and
$P^\theta_{(C)} \to \frac{\partial L}{\partial \dot{\theta}} P^\phi_{(C)} + P^\theta_{(\alpha)}$,
where $\alpha$ is the angle \eqref{alpha} in the plane of motion. 
Alternatively,
these components can be derived from expression \eqref{P.C}
by means of the chain rule:
\begin{subequations}\label{gen.P.C}
\begin{align}
  P^t_{(C)} & =
  -\frac{r}{r-2M} \Big(
  \frac{\partial\tilde C}{\partial E} \frac{\partial E}{\partial\dot{t}}
 + \frac{\partial\tilde C}{\partial N} \frac{\partial N}{\partial\dot{t}}
  \Big) , 
  \\
  P^r_{(C)} & =
  \frac{r-2M}{r} \frac{\partial\tilde C}{\partial N} \frac{\partial N}{\partial\dot{r}} , 
  \\
  P^\phi_{(C)} & =
  \frac{1}{r^2\sin^2\theta} \Big( 
  \frac{\partial\tilde C}{\partial L} \frac{\partial L}{\partial\dot{\phi}}
  + \frac{\partial\tilde C}{\partial L^x} \frac{\partial L^x}{\partial\dot{\phi}}
  + \frac{\partial\tilde C}{\partial L^y} \frac{\partial L^y}{\partial\dot{\phi}}
  + \frac{\partial\tilde C}{\partial N} \frac{\partial N}{\partial\dot{\phi}}
  \Big) , 
  \\
  P^\theta_{(C)} & =
  \frac{1}{r^2} \Big( 
  \frac{\partial\tilde C}{\partial L} \frac{\partial L}{\partial\dot{\theta}}
  + \frac{\partial\tilde C}{\partial L^x} \frac{\partial L^x}{\partial\dot{\theta}}
  + \frac{\partial\tilde C}{\partial L^y} \frac{\partial L^y}{\partial\dot{\theta}}
  + \frac{\partial\tilde C}{\partial N} \frac{\partial N}{\partial\dot{\theta}}
  \Big) 
\end{align}
\end{subequations}
with  
\begin{equation}\label{gen.tilC}
  \tilde C(\lambda,t,r,\phi,\theta,L^x,L^y,L,E,N) = C
\end{equation}
where the geodesic variables $\frac{dt}{d\tau}$, $\frac{dr}{d\tau}$, $\frac{d\phi}{d\tau}$, $\frac{d\theta}{d\tau}$
are expressed in terms of $L^x$, $L^y$, $L$, $E$ in the given conserved quantity $C$. 
Detailed evaluation of these variational symmetry components for $C=\Phi,T,\T$
are not needed subsequently and hence they will be omitted.

For the LRL direction angle $(\A^x,\A^y,\A^z)$,
the variational symmetries are given in terms of
$\X_{(\Phi)}$, $\X_{(L^i)}$ through applying the chain rule \eqref{prX.chainrule}
to expressions \eqref{Ax.Ay.Az},
yielding
\begin{equation}
  \X_{(A^i)} =
  \frac{\partial A^i}{\partial \Phi} \X_{(\Phi)}
+ \frac{\partial A^i}{\partial L^x} \X_{(L^x)} 
+ \frac{\partial A^i}{\partial L^y} \X_{(L^y)}
+ \frac{\partial A^i}{\partial L^z} \X_{(L^z)} ,
\quad
i = x, y, z. 
\end{equation}

\subsection{Action of variational symmetries on conserved quantities}\label{subsec:symm.actions}

Each variational symmetry $\X_{(C)}$ arising from
$E$, $L^x$, $L^y$, $L^z$, $\Phi$, $T$, $\T$, $\A^x$, $\A^y$, $\A^z$ 
gives a prolonged linear differential operator \eqref{prX}
in terms of the geodesic Lagrangian variables. 
These operators $\pr\X_{(C)}$ act on the function space of
all locally conserved quantities of the timelike geodesic equations 
such that each pair of locally conserved quantities $C_1$ and $C_2$
produces another locally conserved quantity $C_3 = \pr\X_{(C_1)} C_2$. 
This action satisfies the general property \eqref{prX.C1.C2}
that $C_3$ is antisymmetric in $C_1$ and $C_2$. 

The aim here will be to derive the symmetry actions involving
the set of locally conserved quantities in Theorem~\ref{thm:cons.quantities}. 
This will be done by use of the action of the $SO(3)$ group
generated by the rotational Killing vectors \eqref{rotation.KVs} 
which will be applied to the symmetry actions derived in the equatorial case, 
similarly to the derivation of $(\A^x,\A^y,\A^z)$ in subsection~\ref{subsec:cons.quantities}.

First,
start from the equatorial symmetry actions
\eqref{L.E.actions.L.E}, \eqref{L.E.actions.Phi.T}, \eqref{Phi.T.actions.L.E}
involving $E$, $L$, $\Phi$.
They can be expressed as
\begin{gather}
  \pr\X_{(E)} (0,0,L) = -\pr\X_{(0,0,L)} E  = 0,
\\
  \pr\X_{(\Phi)} E = -\pr\X_{(E)} \Phi = 0,
  \quad
  \pr\X_{(\Phi)} (0,0,L) = -\pr\X_{(0,0,L)} \Phi = -(0,0,1) . 
\label{Phi.actions.L.E}
\end{gather}
Now apply the $SO(3)$ rotations \eqref{azimuth.rotation}--\eqref{polar.rotation}
which carry $(0,0,L)$ to $(L^x,L^y,L^z)$. 
Since $E$ and $L$ are rotationally invariant,
this yields
\begin{equation}\label{gen.L.E.action.L.E}
  \pr\X_{(L^x,L^y,L^z)} E = -\pr\X_{(E)} (L^x,L^y,L^z) = 0
\end{equation}
and
\begin{gather}
  \pr\X_{(\Phi)} E = - \pr\X_{(E)} \Phi = 0
\label{gen.Phi.action.E}
  \\
  \pr\X_{(\Phi)} (L^x,L^y,L^z) = - \pr\X_{(L^x,L^y,L^z)} \Phi = -L^{-1} (L^x,L^y,L^z), 
\label{gen.Phi.action.L}
\end{gather}
where $E$, $L$, and $\Phi$ are the non-equatorial conserved quantities \eqref{E}, \eqref{L}, and \eqref{gen.Phi}. 

Second, 
the analogous symmetry actions involving $\pr\X_{(\A^x,\A^y,\A^z)}$
can be derived from expressions \eqref{Phi.actions.L.E} 
by starting with $(\cos\Phi,\sin\Phi,0)$
and applying the chain rule \eqref{prX.chainrule}: 
\begin{gather}
  \pr\X_{(E)} (\cos\Phi,\sin\Phi,0)  = (\pr\X_{(E)}\Phi)(-\sin\Phi,\cos\Phi,0) 
  = (0,0,0) , 
\\
  \pr\X_{(L)} (\cos\Phi,\sin\Phi,0)  = (\pr\X_{(L)}\Phi)(-\sin\Phi,\cos\Phi,0)
  = (0,0,1) \times (\cos\Phi,\sin\Phi,0) , 
 \label{L.action.Phi}
\end{gather}
where the last equality uses the vector cross product. 
As the $SO(3)$ rotations \eqref{azimuth.rotation}--\eqref{polar.rotation}
simultaneously carry 
$(0,0,L)$ and $(\cos\Phi,\sin\Phi,0)$ to $(L^x,L^y,L^z)$ and $(\A^x,\A^y,\A^z)$,
while preserving the cross product, 
this yields
\begin{equation}\label{gen.E.action.Ax.Ay.Az}
  \pr\X_{(E)} (\A^x,\A^y,\A^z)  = 0
\end{equation}
and
\begin{equation}\label{gen.Lx.Ly.Lz.action.Ax.Ay.Az}
  \begin{aligned}
    & \pr\X_{(L^x,L^y,L^z)} (\A^x,\A^y,\A^z) 
    \\&\quad
    = \big( (1,0,0) \times (\A^x,\A^y,\A^z), (0,1,0) \times (\A^x,\A^y,\A^z), (0,0,1) \times (\A^x,\A^y,\A^z) \big)
    \\&\quad
    = 
    \begin{pmatrix}
    0 & -\A^z & \A^y \\
    \A^z & 0 & -\A^x \\
    -\A^y &  \A^x & 0
    \end{pmatrix} . 
  \end{aligned}
\end{equation}
Property \eqref{prX.C1.C2} then gives
\begin{equation}\label{gen.Ax.Ay.Az.action.E}
  \pr\X_{(\A^x,\A^y,\A^z)} E  = 0 
\end{equation}
and
\begin{equation}\label{gen.Ax.Ay.Az.action.Lx.Ly.Lz}
  \pr\X_{(\A^x,\A^y,\A^z) } (L^x,L^y,L^z)
  = - 
    \begin{pmatrix}
    0 & -\A^z & \A^y \\
    \A^z & 0 & -\A^x \\
    -\A^y &  \A^x & 0
    \end{pmatrix} .
\end{equation}
Similarly, 
starting from 
\begin{equation}
  \pr\X_{(\cos\Phi,\sin\Phi)} (\cos\Phi,\sin\Phi,0)  = 0
\end{equation}
which holds by the chain rule, 
and again applying the same rotations leads to the symmetry action 
\begin{equation}\label{gen.Ax.Ay.Az.action.Ax.Ay.Az}
  \pr\X_{(\A^x,\A^y,\A^z) } (\A^x,\A^y,\A^z) =0 .
\end{equation}
Likewise,
\begin{equation}\label{gen.Phi.action.Ax.Ay.Az}
  \pr\X_{(\Phi) } (\A^x,\A^y,\A^z) =0 .
\end{equation}

The symmetry action of $\pr\X_{(L^i)}$ on $L^j$, for $i,j=x,y,z$, is given by 
\begin{equation}\label{gen.LxLyLz.action.Lx,LyLz}
  \pr\X_{(L^x,L^y,L^z)} (L^x,L^y,L^z)   =
    \begin{pmatrix}
    0 & -L^z & L^y \\
    L^z & 0 & -L^x \\
    -L^y &  L^x & 0
  \end{pmatrix} ,
\end{equation}
which follows from the fact that $\pr\X_{(L^x)}$, $\pr\X_{(L^y)}$, $\pr\X_{(L^z)}$
act in $\Rnum^3$ as the generators of rotations in the directions $x,y,z$ respectively.
In particular,
this symmetry action \eqref{gen.LxLyLz.action.Lx,LyLz} coincides with
the Killing vectors $\xi_{(i)}$, $i=x,y,z$, acting on $(L^x,L^y,L^z)$
viewed as a triplet function on timelike geodesics. 

It is worth noting that the symmetry actions \eqref{gen.LxLyLz.action.Lx,LyLz} and \eqref{gen.Phi.action.L}
can be used together with expressions \eqref{Ax.Ay.Az}
to provide an alternative derivation of the symmetry action \eqref{gen.Lx.Ly.Lz.action.Ax.Ay.Az},
which involves a direct computation utilizing the chain rule 
\begin{equation}
\pr\X_{(L^i)} \A^j =
\frac{\partial\A^j}{\partial \Phi} \pr\X_{(L^i)}\Phi
+ \frac{\partial\A^j}{\partial L^k} \pr\X_{(L^i)} L^k . 
\end{equation}
The following identity, which expresses the orthogonality of 
$(L^x,L^y,L^z)$ and $(\sin\theta\cos\phi,\sin\theta\sin\phi,\cos\theta)$, 
\begin{equation}\label{Lx.Ly.Lz.identity}
  L^z\cot\theta  + L^x\cos\phi + L^y\sin\phi =0
\end{equation}
is useful for these computations. 

Next,
the symmetry actions by $(L^x,L^y,L^z)$ and $E$ on $T$ and $\T$
can be derived from the equatorial symmetry actions \eqref{L.E.actions.Phi.T}--\eqref{L.E.actions.Tau}
similarly to the result \eqref{gen.E.action.Ax.Ay.Az},
yielding
\begin{gather}
\pr\X_{(L^x,L^y,L^z)}T =0,
\quad
\pr\X_{(E)}T =-1, 
\label{gen.LxLyLz.E.actions.T}
\\
\pr\X_{(L^x,L^y,L^z)}\T =- (L^x,L^y,L^z), 
\quad
\pr\X_{(E)}\T =-E  . 
\label{gen.LxLyLz.E.actions.Tau}
\end{gather}

Finally,
consider the equatorial symmetry actions \eqref{T.Phi.actions}--\eqref{Tau.T.actions}
among $\Phi$, $T$, $\T$,
which are given by the functions \eqref{C.LRL} of $L$ and $E$.
These symmetry actions carry over to the non-equatorial case 
with $E$ and $L$ replaced by
the energy \eqref{E} and the scalar angular momentum \eqref{L}. 
This gives, similarly to the steps \eqref{L.action.Phi}, 
\begin{align}
  & \pr\X_{(T)} (\A^x,\A^y,\A^z)  
  = L^{-1} C_{T,\Phi}\, (L^x,L^y,L^z)\times (\A^x,\A^y,\A^z) ,  
\label{gen.T.Ax.Ay.Az.actions}\\
& \pr\X_{(\T)} (\A^x,\A^y,\A^z)  
  = L^{-1} C_{\T,\Phi}\, (L^x,L^y,L^z)\times (\A^x,\A^y,\A^z) . 
\label{gen.Tau.Ax.Ay.Az.actions}
\end{align}

The preceding results are summarized as follows. 

\begin{prop}\label{prop:symm.actions}
For non-equatorial timelike geodesics,
the ten conserved quantities $E$, $(L^x,L^y,L^z)$, $\Phi$, $(\A^x,\A^y,\A^z)$, $T$ and $\T$
correspond to variational symmetries whose action on these quantities 
is given by expressions 
\eqref{gen.L.E.action.L.E}--\eqref{gen.Phi.action.L},
\eqref{gen.E.action.Ax.Ay.Az}--\eqref{gen.Ax.Ay.Az.action.Lx.Ly.Lz}, 
\eqref{gen.Ax.Ay.Az.action.Ax.Ay.Az}--\eqref{gen.LxLyLz.action.Lx,LyLz},
\eqref{gen.LxLyLz.E.actions.T}--\eqref{gen.Tau.Ax.Ay.Az.actions},
and \eqref{T.Phi.actions}--\eqref{C.LRL.cp}.
Only seven of the ten quantities are independent, since
$\Phi$ is determined by $(\A^x,\A^y,\A^z)$
which has unit norm and is orthogonal to $(L^x,L^y,L^z)$
due to relations \eqref{A.rels}. 
\end{prop}

\section{Noether symmetry group}\label{sec:group}

The non-equatorial variational symmetries corresponding to
the energy $E$, angular momentum $(L^x,L^y,L^z)$,
LRL angle $\Phi$, LRL time $T$ and proper time $\T$
generate the full Noether symmetry group $\G$ under compositions, 
modulo the group generated by re-parameterization symmetries
$f(L^x,L^y,L^z,E)\pr\X_\shft$
(involving smooth functions $f$).
The Lie algebra structure of $\G$ as a local Lie group will now be stated. 

To begin, a direct computation yields the commutators of the variational symmetries
arising from $E$, $(L^x,L^y,L^z)$, $\Phi$, $T$, and $\T$.
The only change in the commutators compared to the equatorial case consists of 
\begin{gather}
[\pr\X_{(L^i)},\pr\X_{(E)}] =0,
\quad
[\pr\X_{(L^i)},\pr\X_{(T)}] =0,
\\
[\pr\X_{(L^i)},\pr\X_{(\Phi)}] =L^{-1}L^i\pr\X_\shft,
\quad
[\pr\X_{(L^i)},\pr\X_{(\T)}] =-L^i\pr\X_\shft , 
\end{gather}
and
\begin{equation}
[\pr\X_{(L^i)},\pr\X_{(L^j)}] = -\epsilon_{ij}{}^k \pr\X_{(L^k)} ,
\end{equation}
where, latin indices run $i=x,y,z$; 
$\epsilon_{ij}{}^{k}$ is the Levi-Civita symbol
(namely, $\epsilon^{ij}{}_{k}=\pm 1$ for $(i,j,k)$
being an even and odd permutation of $(x,y,z)$ respectively);
and summation over repeated indices is assumed. 
In particular,
in the equatorial commutators \eqref{comm.0}--\eqref{comm.Cs},
all of the ones that involve $\pr\X_{(E)}$, $\pr\X_{(\Phi)}$, $\pr\X_{(T)}$, $\pr\X_{(\T)}$
remain the same other than replacing $E$, $L$, and $\Phi$
by the non-equatorial expressions \eqref{E}, \eqref{L}, and \eqref{gen.Phi}.

As a consequence, 
the Lie algebra of $\G$ has a somewhat more complicated structure
compared to equatorial case in Theorem~\ref{thm:liealgebra}.

\begin{thm}\label{thm:SO3.liealgebra}
The structure of the Lie algebra $\alg{g}$ of
the non-equatorial Noether symmetry group $\G$ 
depends on whether the LRL quantities $\Phi$, $T$, $\T$ 
are formulated using a turning point, a centripetal point, or a horizon-crossing point.
Let
$\alg{f} = \spans(f(L^x,L^y,L^z,E)\pr\X_\shft)$ generated by (smooth) functions $f$,
and 
$\alg{k} = \spans(\pr\X_{(f(L^x,L^y,L^z,E))})$ generated by (smooth) functions $f$,
which define infinite-dimensional Lie algebras,
where $\alg{f}$ is abelian
and $\alg{k}$ has non-zero Lie bracket
\begin{equation}
  [\pr\X_{(f(L^i,E))},\pr\X_{(g(L^j,E))}]
  = \pr\X_{(h(L^k,E))},
  \quad
  h = \epsilon^{ij}{}_k f_{L^i} g_{L^j} L^k
\end{equation}
\\
(i) In the case of a turning point or a horizon-crossing point,
the Lie algebra is a semi-direct sum of subalgebras
\begin{equation}
 \alg{g} = \alg{h} \ltimes(\alg{k}\oplus\alg{f})
\end{equation}
where $\alg{f}$ is an ideal;
$\alg{k}$ is a non-abelian subalgebra; 
$\alg{h} = \spans(\pr\X_{(\Phi)},\pr\X_{(\T)},\pr\X_{(T)})$ is a 3-dimensional subalgebra
with non-zero Lie brackets
\begin{equation}
  [\pr\X_{(\T)},\pr\X_{(\Phi)}] = -\pr\X_{(\Phi)},
  \quad
 [\pr\X_{(\T)},\pr\X_{(T)}] = -\pr\X_{(T)} ; 
\end{equation}
and $\alg{h}$ acts on $\alg{k}\oplus\alg{f}$ via the Lie brackets
\begin{subequations}\label{gen.comm.not0}
  \begin{align}
  [\pr\X_{(\Phi)},\pr\X_{(f(L^i,E))}]
  & = - L^{-1} L^i\big( \pr\X_{(f_{L^i})} + f_{L^i}\pr\X_\shft \big) , 
\\
  [\pr\X_{(T)},\pr\X_{(f(L^i,E))}]
  & = \pr\X_{(f_{E})} + f_{E}\pr\X_\shft , 
\\
  [\pr\X_{(\T)},\pr\X_{(f(L^i,E))}]
   & = L^i\pr\X_{(f_{L^i})} + E\pr\X_{(f_{E})} + (L^i f_{L^i} + E f_{E}) \pr\X_\shft
\end{align}
\end{subequations}    
\\
(ii) In the case of a centripetal point,
the Lie algebra is a direct sum of vector spaces
\begin{equation}
 \alg{g} = \Rnum^3 \oplus \alg{k}\oplus\alg{f}
\end{equation}
where $\alg{f}$ is an ideal;
$\alg{k}$ is a non-abelian subalgebra; 
$\Rnum^3 = \spans(\pr\X_{(\Phi)},\pr\X_{(\T)},\pr\X_{(T)})$ is a subspace
with the Lie brackets \eqref{comm.Cs},
and it acts on $\alg{k}\oplus\alg{f}$ via the Lie brackets \eqref{gen.comm.not0}. 
\end{thm}

\subsection{Symmetry transformation groups generated by LRL quantities}
From Proposition~\ref{prop:symm.group},
each variational symmetry produces a 
one-dimensional local Lie group of symmetry transformations,
whose compositions generate the Noether symmetry group.
The local Lie groups arising from the LRL quantities
consist of dynamical symmetry transformations.
Their explicit form will now be worked out
by the same method used for the equatorial case in section~\ref{sec:transformations}. 

In outline,
the dynamical symmetry transformations act in the extended space of variables
$(\tau,t,r,\phi,\theta,\sigma,v,\omega,\varpi)$,
with $\sigma =\frac{dt}{d\tau}$, $v=\frac{dr}{d\tau}$, $\omega=\frac{d\phi}{d\tau}$, $\varpi=\frac{d\theta}{d\tau}$.
Each transformation group will have the form
\begin{equation}\label{gen.Y.group}
  (\tau,t,r,\phi,\theta,\sigma,v,\omega,\varpi) \to
  (\tau^\dagger,t^\dagger,r^\dagger,\phi^\dagger,\theta^\dagger,\sigma^\dagger,v^\dagger,\omega^\dagger,\varpi^\dagger)
  =  \exp\big(\varepsilon\pr\Y_{(C)}\big)(\tau,t,r,\phi,\theta,\sigma,v,\omega,\varpi) 
\end{equation}
using the equivalent symmetry generator
$\pr\Y_{(C)}= \chi N^{-1} D_\lambda +\pr\X_{(C)}$
which acts the same as the variational symmetry $\pr\X_{(C)}$ on all geodesics.
Here $\varepsilon$ is the group parameter,
and $D_\lambda$ is a total derivative operator with respect to the (non-affine) parameter $\lambda$,
with $(\ddot{t},\ddot{r},\ddot{\phi},\ddot{\theta})$
satisfying the Euler-Lagrange equations \eqref{gen.geodesic.eqns}
of the geodesic Lagrangian. 

For each variational symmetry corresponding to a conserved quantity $C$, 
the gauge choice $\pr\Y_{(C)} r =0$ is imposed,
which determines $\chi$. 
As a preliminary step, 
expressions \eqref{alpha}--\eqref{absv}, \eqref{gen.T}--\eqref{gen.Tau}
can be inverted to obtain the relations
\begin{subequations}\label{gen.dynvars}
\begin{gather}
  \tau = \T +\sgn(v) \int_{r_0}^{r} \frac{dy}{|v(y,L,E)|},
  \quad
   t = T +  \sgn(v) E \int_{r_0}^{r} \frac{y\, dy}{(y-2M) |v(y,L,E)|},
\\
  \sigma = (1 - 2M/r)^{-1} E,
  \quad
  v = \sgn(v)\sqrt{E^2-(1- 2M/r)(1+L^2/r^2)},
\\
\varpi = (L^y\cos\phi - L^x\sin\phi)/r^2,
\quad
\omega = ( (L^x\cos\phi + L^y\sin\phi)^2 +(L^z)^2 )/(r^2L^z), 
\\
\tan\theta = - L^z/(L^x\cos\phi + L^y\sin\phi),
\\
\quad
\tan\phi = ( L L^z\tan\alpha\, - L^x L^y)/((L^y)^2 + (L^z)^2), 
\\
\alpha =  \Phi + \sgn(v) L \int_{r_0}^{r} \frac{dy}{y^2|v(y,L,E)|},
\quad
L = L^z \sqrt{1 + (L^x/L^z)^2 + (L^y/L^z)^2}
\end{gather}
\end{subequations}
from which the infinitesimal symmetry transformations 
are determined from the expressions shown in Proposition~\ref{prop:symm.actions}
for $\pr\X_{(C)}$ acting on $E$, $(L^x,L^y,L^z)$, $\Phi$, $T$, $\T$. 

The following result extends Theorems~\ref{thm:Phi.group}--\ref{thm:Tau.group}
to the non-equatorial case. 

\begin{thm}\label{thm:symm.group}
(i) 
The LRL angular quantity $\Phi$ generates
a dynamical symmetry transformation group, with parameter $\varepsilon$,
given by equations \eqref{gen.dynvars}
with 
\begin{equation}
  L^i\to L^i{}^\dagger = (1 - \varepsilon L^{-1}) L^i,
  \quad
  i= x,y,z
\end{equation}
and, in the case when $r_0$ is a centripetal point, 
\begin{equation}
T \to T^\dagger = T - \int_{L}^{L^\dagger} C_{T,\Phi}\,dL, 
\quad
\T \to \T^\dagger = \T - \int_{L}^{L^\dagger} C_{\T,\Phi}\,dL
\end{equation}
and
\begin{equation}
  r_0 \to r_0^\dagger=r_0(L^\dagger,E)
\end{equation}
with $L^\dagger = L -\varepsilon$, 
where $C_{T,\Phi}$ and $C_{\T,\Phi}$ are given by expressions \eqref{C.LRL.cp}.
\\
(ii) 
The LRL temporal quantity $T$ generates
a dynamical symmetry transformation group, with parameter $\varepsilon$,
 given by equations \eqref{gen.dynvars}
with 
\begin{equation}
  E\to E^\dagger = E+\varepsilon
\end{equation}
and, in the case when $r_0$ is a centripetal point, 
\begin{equation}
\Phi \to \Phi^\dagger = \Phi + \int_{E}^{E^\dagger} C_{T,\Phi}\,dE, 
\quad
\T \to \T^\dagger = \T -\int_{E}^{E^\dagger} C_{\T,T}\,dE
\end{equation}
and
\begin{equation}
  r_0 \to r_0^\dagger=r_0(L,E^\dagger)
\end{equation}
where $C_{T,\Phi}$ and $C_{\T,\T}$ are given by expressions \eqref{C.LRL.cp}
in terms of $E$ and $L$. 
\\
(iii) 
The LRL proper-time quantity $\T$ generates
a dynamical symmetry transformation group, with parameter $\varepsilon$,
given by equations \eqref{gen.dynvars}
with 
\begin{equation}
  E\to E^\dagger = \exp(\varepsilon) E,
  \quad
  L^i\to   L^i{}^\dagger = \exp(\varepsilon) L^i,
  \quad
  i= x,y,z  
\end{equation}
and, in the case when $r_0$ is a centripetal point,
\begin{subequations}
\begin{align}
\Phi \to   \Phi^\dagger & = \Phi +\int_{\sqrt{|LE|}}^{\sqrt{|LE|}^\dagger} C_{\T,\Phi}(\sgn(L) y \sqrt{|L|/|E|},\sgn(E) y \sqrt{|E|/|L|}) \,\frac{dy}{y},
\\
T \to   T^\dagger & = T +\int_{\sqrt{|LE|}}^{\sqrt{|LE|}^\dagger} C_{\T,T}(\sgn(L) y \sqrt{|L|/|E|},\sgn(E) y \sqrt{|E|/|L|}) \,\frac{dy}{y}
\end{align}
\end{subequations}
and
\begin{equation}
  r_0 \to r_0^\dagger=r_0(L^\dagger,E^\dagger)
\end{equation}
where $C_{\T,\Phi}$ and $C_{\T,\T}$ are given by expressions \eqref{C.LRL.cp}
in terms of $E$ and $L$. 
\end{thm}

In each transformation group,
the parameter space $\{\varepsilon\}$ is constrained by the inequalities
stated after Theorems~\ref{thm:Phi.group}--\ref{thm:Tau.group} respectively.
The physical meaning of these symmetry groups is explained by
their action on $E$ and $(L^x,L^y,L^z)$, 
which is analogous to the discussion for equatorial geodesics
in sections~\ref{sec:Phi.transformations}--\ref{sec:Tau.transformations}.

\subsection{Local Lie group structure}

In addition to the dynamical symmetry transformations
generated from the LRL quantities $\Phi$, $T$, $\T$
in Theorem~\ref{thm:symm.group},
the energy $E$ and angular momenta $(L^x,L^y,L^z)$
generate point symmetry transformations which coincide 
with the isometry groups arising from the time-translation and rotation Killing vectors. 
The latter are Lie groups, $\Rnum$ and $SO(3)$,
whereas the former constitute one-dimensional local Lie groups. 

Composition of the three local Lie groups and the isometry groups
yields the Noether symmetry group modulo the group generated by
re-parameterization symmetries $f(L^x,L^y,L^z,E)\pr\X_\shft$
(involving smooth functions $f$).

\section{Concluding remarks}\label{sec:conclude}

The Noether symmetry group of the timelike non-radial geodesic equations
in Schwarzschild spacetime
has been shown to have a hidden structure consisting of 
three symmetry transformations which come from
corresponding hidden conserved quantities
derived recently in \Ref{Anc.Faz}. 
In particular, the geodesic equations possess
an LRL angle, an LRL Killing-vector time and an LRL proper time,
which are (piecewise) conserved along the geodesics. 
Applying Noether's theorem in reverse to these quantities
yields the new symmetry transformations of the geodesic Lagrangian.

The existence of these hidden symmetries is a novel result
which has previously gone unnoticed in General Relativity. 

In contrast to the Killing vector symmetries,
which represent point transformations
on the dynamical configuration variables $(t,r,\phi,\theta)$ in the geodesic equations,
the new symmetries describe dynamical transformations
which essentially involve the momenta variables
$(\frac{dt}{d\tau},\frac{dr}{d\tau},\frac{d\phi}{d\tau},\frac{d\theta}{d\tau})$
in the geodesic equations.
An alternative setting is provided by the Hamiltonian formulation of the geodesic equations,
where the configuration and momenta variables are on equal footing.
The LRL angle, LRL Killing-vector time, and LRL proper time,
along with the energy and angular momenta,
form a Poisson algebra in this setting. 

The LRL angle can be combined with the conserved angular momenta $(L^x,L^y,L^z)$ 
to obtain a conserved LRL direction vector $(\A^x,\A^y,\A^z)$,
which belongs to the representation space $\Rnum^3$ for the $SO(3)$ rotation group.
This direction vector gives rise to three dynamical symmetry transformations
through Noether's theorem in reverse.
These infinitesimal symmetries mutually commute,
and under rotations they transform as components of a vector.
Moreover, a conserved LRL vector can be defined from the direction vector
and a conserved scalar, 
in analogy to the situation for Newtonian gravity.

It is worth emphasizing that, in contrast to the Killing vector symmetries,
the infinitesimal dynamical symmetries coming from the LRL quantities
do not represent some intrinsic geometrical structure on Schwarzschild spacetime.

In future work,
the explicit form of the symmetry transformations arising from
the LRL direction vector and the LRL vector will be derived,
and the structure of the resulting local Lie group will be determined \cite{Anc.Gol}. 
The method of the present work can be applied directly to find
all hidden conserved quantities and associated hidden symmetries
for any spherically spacetime.

An interesting open problem will be to investigate 
LRL-type (hidden) conserved quantities and symmetries
for the null geodesics in Schwarzschild spacetime.
Hidden conserved quantities and symmetries for timelike and null geodesics
can also be explored for spacetimes with other symmetries, 
especially the Kerr black hole.

\appendix
\section{Solution of the group parameter inequalities}

The solution of the inequalities that define the range of the group parameter
in Theorems~\ref{thm:Phi.group} and~\ref{thm:Tau.group}
will be derived here.

In Theorem~\ref{thm:Phi.group},
the inequalities can be conveniently written in terms of
$p = 1/\bar{L}^\dagger{}^2 =1/(\bar{L} -\bar{\varepsilon})^2$.
Note that
\begin{equation}\label{B.C}
  E^2 - E_\pm^\dagger{}^2
  = \tfrac{2}{27} p^{-1} (B \mp C),
  \quad
  B = (\tfrac{27}{2}E^2 - 9)p  - 1,
  \quad
  C=\sqrt{(1 - 3p)^3} . 
\end{equation}

For a centripetal point,
there are two inequalities to be solved, 
$p\leq \tfrac{1}{3}$ and $B + C\geq 0$.
The latter is equivalent to the union of the polynomial inequalities
\begin{subequations}\label{Phi.group.cp.ineqn}
  \begin{gather}
    p\leq \tfrac{1}{3},
    \quad
    B\geq 0
    \label{Phi.group.cp.ineqn1}
    \\
    p\leq \tfrac{1}{3},
    \quad
    B<0 ,
    \quad
    C^2 \geq B^2
    \label{Phi.group.cp.ineqn2}
  \end{gather}
\end{subequations}
since $C\geq 0$ due to $p\leq \tfrac{1}{3}$.
By means of the factorization 
\begin{equation}
  C^2 -B^2 = -\tfrac{1}{27}p^{-1}\big( p^2 + \tfrac{27}{4}((E^2 -\tfrac{2}{3})^2 -1)p - E^2 + 1 \big) , 
\end{equation}
these polynomial inequalities \eqref{Phi.group.cp.ineqn} 
have the respective solutions
\begin{equation}\label{Phi.group.cp.ineqn1.soln}
  1/a \leq p \leq \tfrac{1}{3} ,
  \quad
  E^2 > \tfrac{8}{9}
\end{equation}
and
\begin{equation}\label{Phi.group.cp.ineqn2.soln}
  \begin{gathered}
  1/\bar{L}_-^2 \leq p < 1/a,
  \quad
  \tfrac{8}{9} < E^2 < 1
  \\
  0 < p < 1/a,
  \quad
  E^2 \geq 1
  \end{gathered}
\end{equation}
where $a = \tfrac{27}{2}(E^2 - \tfrac{2}{3}) >0$
due to $E^2\geq E_-^2 \geq \tfrac{8}{9}$. 
Then the union of \eqref{Phi.group.cp.ineqn1.soln} and \eqref{Phi.group.cp.ineqn2.soln}
consists of
\begin{equation}\label{Phi.group.cp.ineqn.soln}
  \begin{gathered}
  p \leq \tfrac{1}{3},
  \quad
  E^2 \geq 1 ; 
  \\
  1/\bar{L}_-^2 \leq p \leq \tfrac{1}{3}
  \quad
  \tfrac{8}{9} < E^2 < 1 . 
  \end{gathered}
\end{equation}

For a turning point,
the inequalities to be solved are given by
$p>\tfrac{1}{4}$ and $E^2<1$, 
along with $p\leq \tfrac{1}{4}$ and $B-C\leq 0$.
The latter two are equivalent to the union of the polynomial inequalities
\begin{subequations}
  \begin{gather}
    p\leq \tfrac{1}{4},
    \quad
    B\leq 0
    \label{Phi.group.cp.ineqn1}
    \\
    p\leq \tfrac{1}{4},
    \quad
    B>0 ,
    \quad
    C^2 \geq B^2
    \label{Phi.group.cp.ineqn2}
  \end{gather}
\end{subequations}
since $C\geq 0$ due to $p<\tfrac{1}{3}$.
These polynomial inequalities have the respective solutions
\begin{equation}\label{Phi.group.tp.ineqn1.soln}
  p \leq \begin{cases}
    \tfrac{1}{4}, & E^2 \leq \tfrac{26}{27}
    \\
    1/a, & E^2 > \tfrac{26}{27}
    \end{cases}
\end{equation}
and
\begin{equation}\label{Phi.group.tp.ineqn2.soln}
  \begin{gathered}
  1/a <p \leq \tfrac{1}{4},
  \quad
  \tfrac{26}{27} < E^2 \leq 1
  \\
  1/a < p \leq 1/\bar{L}_+^2,
  \quad
  E^2 > 1
  \end{gathered}
\end{equation}
where $a = \tfrac{27}{2}(E^2 - \tfrac{2}{3}) >0$ due to $E^2>\tfrac{26}{27}$.
Taking the union of \eqref{Phi.group.tp.ineqn1.soln} and \eqref{Phi.group.tp.ineqn2.soln}
with 
\begin{equation}\label{Phi.group.tp.ineqn0.soln}
  p>\tfrac{1}{4},
  \quad
  E^2<1
\end{equation}
then yields simply 
\begin{equation}\label{Phi.group.tp.ineqn.soln}
  p \leq 1/\bar{L}_+^2,
  \quad
  E^2 > 1 . 
\end{equation}

Finally, 
the range for $\bar{\varepsilon}$ consists of the interior of the intervals
\eqref{Phi.group.cp.ineqn.soln} and \eqref{Phi.group.tp.ineqn.soln}. 

In Theorem~\ref{thm:Tau.group},
the inequalities have the simplest form in terms of 
$p' = 1/\bar{L}^\dagger{}^2 = 1/(\bar{L}^2e^{2\varepsilon})$.
Similarly to the relation \eqref{B.C},
note that
\begin{equation}\label{B.C}
  E^\dagger{}^2 - E_\pm^\dagger{}^2
  = \tfrac{2}{27} p'{}^{-1} (B' \mp C'),
  \quad
  B' = b^{-2} - 1 - 9p', 
  \quad
  C'=\sqrt{(1 - 3p')^3}
\end{equation}
where $b$ is expression \eqref{b}. 
Unlike in the previous case,
here the cubic polynomial $C'{}^2 - B'{}^2$ does not reduce to a quadratic polynomial. 
Its positive roots, $p'=q>0$, are explicitly given by
\begin{subequations}
  \begin{align}
    q_{\max} & =
    \tfrac{2}{3}\bigg( \sqrt{2b^{-2} + 1} \cos\Big(\tfrac{1}{3}\arctan\Big(\frac{b^{-1}\sqrt{4-b^{-2}}^3}{b^{-4} + 10b^{-2} - 2}\Big) -\tfrac{1}{3}\pi\Big) -1 \bigg) , 
  \\
  q_{\min} & =
      \tfrac{2}{3}\bigg( \sqrt{2b^{-2} + 1} \cos\Big(\tfrac{1}{3}\arctan\Big(\frac{b^{-1}\sqrt{4-b^{-2}}^3}{b^{-4} + 10b^{-2} - 2}\Big) +\tfrac{1}{3}\pi\Big) -1 \bigg) . 
  \end{align}
\end{subequations}

For a centripetal point,
the inequalities to be solved are given by 
$p'\leq \tfrac{1}{3}$ and $B' + C'\geq 0$,
which are equivalent to the union of the polynomial inequalities
\begin{subequations}\label{Tau.group.cp.ineqns}
  \begin{gather}
    p'\leq \tfrac{1}{3},
    \quad
    B'\geq 0
    \label{Tau.group.cp.ineqn1}
    \\
    p'\leq \tfrac{1}{3},
    \quad
    B'<0 ,
    \quad
    C'{}^2 \geq B'{}^2
    \label{Tau.group.cp.ineqn2}
  \end{gather}
\end{subequations}
since $C'\geq 0$ due to $p'\leq \tfrac{1}{3}$.
The respective solutions of these polynomial inequalities \eqref{Tau.group.cp.ineqns}
are given by 
\begin{equation}\label{Tau.group.cp.ineqn1.soln}
  p' \leq \begin{cases} 
    \tfrac{1}{9}(b^{-2}-1), & 1 < b^{-2} < 4 \\
    \tfrac{1}{3}, & 4 \leq b^{-2}
    \end{cases}
\end{equation}
and 
\begin{equation}\label{Tau.group.cp.ineqn2.soln}
  \begin{gathered}
  \tfrac{1}{9}(b^{-2}-1) < p' \leq q_{\max}, 
  \quad
  1 < b^{-2} < 4
  \\
  p' \leq q_{\max},
  \quad
  b^{-2} \leq 1 . 
  \end{gathered}
\end{equation}
Then the union of \eqref{Tau.group.cp.ineqn1.soln} and \eqref{Tau.group.cp.ineqn2.soln}
consists of
\begin{equation}\label{Tau.group.cp.ineqn.soln}
    p' \leq \begin{cases}
      \tfrac{1}{3}, & 4 \leq b^{-2}
      \\
      q_{\max}, &  b^{-2} < 4
      \end{cases} .
\end{equation}

For a turning point,
$p' >\tfrac{1}{4}$ and $p' \geq \tfrac{2}{27} b^{-2}$, 
along with 
$p'\leq \tfrac{1}{4}$ and $B'-C'\leq 0$, 
are the inequalities to be solved.
The latter two are equivalent to the union of the polynomial inequalities
\begin{subequations}\label{Tau.group.tp.ineqn}
  \begin{gather}
    p'\leq \tfrac{1}{4},
    \quad
    B'\leq 0
    \label{Tau.group.tp.ineqn1}
    \\
    p'\leq \tfrac{1}{4},
    \quad
    B'>0 ,
    \quad
    C'{}^2 \geq B'{}^2
    \label{Tau.group.tp.ineqn2}
  \end{gather}
\end{subequations}
since $C'\geq 0$ due to $p'<\tfrac{1}{3}$.
The respective solutions of these inequalities \eqref{Tau.group.tp.ineqn}
are given by 
\begin{equation}\label{Tau.group.tp.ineqn1.soln}
  \begin{gathered}
  p' \leq \tfrac{1}{4}, 
  \quad
  b^{-2} \leq 1
  \\
  \tfrac{1}{9}(b^{-2}-1) < p' \leq \tfrac{1}{4}, 
  \quad
  1<b^{-2} \leq \tfrac{13}{4}
  \end{gathered}
\end{equation}
and
\begin{equation}\label{Tau.group.tp.ineqn2.soln}
  \begin{gathered}
  q_{\min} \leq p' <\begin{cases}
  \tfrac{1}{9}(b^{-2}-1), &   2 < b^{-2} \leq \tfrac{13}{4}
  \\
  \tfrac{1}{4}, & \tfrac{13}{4}  < b^{-2} \leq \tfrac{27}{8}
  \end{cases},
\\
  p' <\tfrac{1}{9}(b^{-2}-1), 
  \quad
  1 < b^{-2} \leq 2 . 
 \end{gathered}
\end{equation}
Taking the union of \eqref{Tau.group.tp.ineqn1.soln} and \eqref{Tau.group.tp.ineqn2.soln}
with 
\begin{equation}\label{Tau.group.tp.ineqn0.soln}
  p'> \begin{cases}
    \tfrac{1}{4}, & b^{-2} < \tfrac{27}{8}
    \\
    \tfrac{2}{27} b^{-2}, & b^{-2} \geq \tfrac{27}{8}
    \end{cases}
\end{equation}
yields simply
\begin{equation}\label{Tau.group.tp.ineqn.soln}
  p' > \begin{cases}
    0, & b^{-2} \leq 2
    \\
    q_{\min} , &   2 < b^{-2} \leq \tfrac{27}{8}
    \\
    \tfrac{2}{27} b^{-2}, & b^{-2} > \tfrac{27}{8}
    \end{cases} .
\end{equation}

Finally, 
the range for $\bar{\varepsilon}$ consists of the interior of the intervals
\eqref{Tau.group.cp.ineqn.soln} and \eqref{Tau.group.tp.ineqn.soln}.


\begin{thebibliography}{99}

\bibitem{Bir}
G.D. Birkhoff,
{\it Relativity and Modern Physics},
(Harvard University Press) 1923.

\bibitem{Mis.Tho.Whe}
C.W. Misner. K.S. Thorne, J.A. Wheeler,
{\it Gravitation}
(W.H. Freeman and Co.) 1973.

\bibitem{Wal}
R.M. Wald,
{\it General Relativity}
(The University of Chicago Press) 1984.

\bibitem{Hag}
Y. Hagihara, 
Theory of the relativistic trajectories in a gravitational field of Schwarzschild, 
Jpn. J. Astron. Geophys. 8 (1931), 67--175.

\bibitem{Dar}
C.G. Darwin, 
The gravity field of a particle, 
Proc. Roy. Soc. A 249 (1959), 180--194; 
The gravity field of a particle II, Proc. Roy. Soc. A 263 (1961), 39--50. 
  
\bibitem{Cha}
S. Chandrasekhar,
{\it The Mathematical Theory of Black Holes}
(Oxford University Press) 1983.


\bibitem{Kat.Lev}
G. Katzin and J. Levine,
Applications of Lie derivatives to symmetries, geodesic mappings, and first integrals in Riemannian spaces, 
Colloquium Math. 26 (1972), 21--38. 

\bibitem{Pri.Cra}
G.E. Prince and M. Crampin,
Projective Differential Geometry and Geodesic Conservation Laws in General Relativity. I: Projective Actions,
Gen. Rel. Grav. 16 (1984), 921--942;
ibid.,
Projective Differential Geometry and Geodesic Conservation Laws in General Relativity, II:
Conservation Laws, 
Gen. Rel. Grav. 16 (1984), 1063--1075.

\bibitem{Bok.Kar.Kas.Zam}
A.H. Bokhari,  A.H. Kara, A.R. Kashif, F.D. Zaman,
Noether Symmetries Versus Killing Vectors and Isometries of Spacetimes,
Internat. J. Theor. Phys. 45(6) (2006), 1029--1039.

\bibitem{Hus.Mah.Qad}
I. Hussain, F.M. Mahomed, A. Qadir,
Approximate partial Noether operators of the Schwarzschild spacetime,
J. Nonlin. Math. Phys. 17(1) (2010), 13--25.

\bibitem{Tsa.Pal}
M. Tsamparlis, A. Paliathanasis,
Lie and Noether symmetries of geodesic equations and collineations,
Gen. Relativ. Grav. 42 (2010), 2957--2980.

\bibitem{Joh.Mah.Kar.Bok.Zam}
A.G. Johnpillai, F.M. Mahomed, A.H. Kara, A.H. Bokhari, F.D. Zaman,
Classification of static spherically symmetric static spacetimes by Noether symmetries,
Internat. J. Theor. Phys.  52 (2013), 3534--3542. 

\bibitem{Ali.Fer.Ali}
F. Ali, T. Feroze, S. Ali, 
Complete classification of spherically symmetric static space-times via Noether symmetries,
Theor. Math. Phys. 184 (2015), 973--985.

\bibitem{Jam.Fer}
B. Jamil, T. Feroze,
Conservation laws corresponding to the Noether symmetries of the geodetic Lagrangian in spherically symmetric spacetimes,
Internat. J. Mod. Phys. D 26(5) (2017),  1741006.

\bibitem{Rez-Hej}
S. Reza Hejazi,
Lie point symmetries, Hamiltonian equations, and conservation laws of the geodesics
on a Schwarzschild black hole, 
Kragujevac Journal of Mathematics 42(3) (2018), 453--475.

\bibitem{Mau.Tor}
D. Maughan, C. Torre, 
Affine symmetry, geodesics, and homogeneous spacetimes,
Gen. Relativ. Grav. 50 (2018), 102. 

\bibitem{Cie.Mac}
A. Cieslik and P. Mach,
Revisiting timelike and null geodesics in the Schwarzschild spacetime: general
expressions in terms of Weierstrass elliptic functions. 
Class. Quantum Grav. 39 (2022), 225003.



\bibitem{Anc.Faz}
S.C. Anco, J. Fazio, 
Analogue of a Laplace-Runge-Lenz vector for particle orbits (timelike geodesics)
 in Schwarzschild spacetime, 
 J. Math. Phys. 64 (2023), 082501.

\bibitem{Gol.Poo.Saf-book}
H. Goldstein, C. Poole, J. Safko, 
{\it Classical Mechanics (3rd ed.)}
(Addison Wesley) 2000.

\bibitem{Cor-book}
B. Cordani, 
{\it The Kepler Problem}
(Birkhaeuser) 2003.

 \bibitem{Anc.Mea.Pas}
S.C. Anco, T. Meadows, V. Pascuzzi,
Some new aspects of first integrals and symmetries for central force dynamics,
J. Math. Phys. 57 (2016), 062901.

\bibitem{Fra}
D.M. Fradkin,
Existence of the dynamic symmetries $O_4$ and $SU(3)$ for all classical central potential problems, 
Prog. Theor. Phys. 37 (1967), 798--812.

\bibitem{Muk}
N. Mukanda,
Dynamical symmetries and classical mechanics,
Phys. Rev. 155 (1967), 1383--1386. 

\bibitem{Per}
A. Peres,
A classical constant of motion with discontinuities,
J. Phys. A: Math. Gen. 12 (1979), 1711--1713.

\bibitem{Olv-book}
P.J. Olver,
{\it Applications of Lie Groups to Differential Equations}
(Springer, New York) 1986.

\bibitem{BA-book}
G. Bluman and S.C. Anco,
{\it Symmetry and Integration Methods for Differential Equations},
Applied Math. Sci. Volume 154
(Springer, New York) 2002.

\bibitem{Ser}
J.P. Serre,
{\it Lie algebras and Lie groups}, 
Springer 1964. 

\bibitem{Anc.2026}
S.C. Anco,
A hybrid Lagrangian-Hamiltonian framework and its application to conserved integrals and symmetry groups,
arXiv:2603.03487.

\bibitem{Whi.Wat-book}
E.T. Whittaker, G.N. Watson,
{\it A Course of Modern Analysis},
(Cambridge University Press) 1915.

\bibitem{Arn-book}
V.I. Arnold,
{\it Mathematical Methods of Classical Mechanics},
Graduate Texts in Mathematics Vol. 50 (2nd ed.)
(Springer) 1989.

\bibitem{Ban.Itz}
M. Bander, C. Itzykson,
Group theory and the hydrogen atom (I),
Rev. Mod. Phys. 38(2) (1966), 330--345.

\bibitem{Anc.Gol.2026}
S.C. Anco, M. Bashmani Moghadam,
Symmetry transformation group arising from the Laplace--Runge--Lenz vector,
Phys. Lett. A 591 (2026), 131926.

\bibitem{Cho}
T.L. Chow,
{\it Classical Mechanics (3rd ed.)}
(CRC Press) 2024.

\bibitem{SnoWin}
L. Snobl, P. Winternitz,
{\it Classification and Identification of Lie algebras},
CRM Monograph Series Vol. 33
(AMS and CRM) 2014.

\bibitem{Anc.Gol}
S.C. Anco, M. Bashmani Moghadam,
A conserved Laplace--Runge--Lenz vector and associated dynamical symmetries
for timelike geodesics in Schwarzschild spacetime. 
(In progress). 




\end{thebibliography}
\end{document}